\documentstyle[12pt,psfig,titlepage]{article}

\title{\vspace*{-4cm}
{\bf The Dirac-Brueckner Approach}\footnote{Prepared for:
{\it Open Problems in Nuclear Matter}, M. Baldo, ed.\ (World
Scientific, Singapore, to be published), Chapter~3.}}
\author{R. Brockmann\\
\it Department of Physics, University of Mainz,\\
\it D-55099 Mainz, Germany\\
\rm and\\
R. Machleidt\\
  \it Department of Physics, University of Idaho,\\
  \it Moscow, ID 83843, U.S.A.}
\date{\today}

\begin{document}

\pagestyle{empty}
\maketitle

\newpage

\tableofcontents

\newpage

\pagestyle{plain}
\section{Introduction}
\pagenumbering{arabic}

One of the most fundamental challenges pervading theoretical nuclear
physics since half a century is to understand the properties of
nuclei in terms of the underlying interactions between the
constituents.

Historically, the first attempt was made by Heisenberg's student
H. Euler who calculated the properties of nuclear matter in second
order perturbation theory \cite{Eu}
assuming nucleons interacting via a two-body potential
of Gaussian shape. When the singular nature
 of the nuclear potential at short distances
('hard core') was realized, it
became apparent that conventional perturbation theory
is inadequate.
Special many-body methods had to be worked out.
Brueckner and coworkers \cite{BLM} initiated a method
which was further developed by
Bethe \cite{Bet} and Goldstone \cite{Gol57}.
Alternatively,
Jastrow \cite{Jas} suggested to take
a variational approach to the nuclear many-body problem.

In the 1960's, substantial advances in the physical understanding
of Brueckner theory were made due to the work by Bethe and coworkers
(see e.~g.\ the review by Day \cite{Day67}). Systematic calculations
of the properties of nuclear matter applying Brueckner theory
started in the late 1960's and continued through the 1970's
[7-9] (see Ref.~\cite{Mac89} for a more recent summary).

The results obtained using a variety of nucleon-nucleon (NN)
potentials show a systematic behaviour: The predictions
for nuclear matter saturation are located along a band
which does not meet the empirical area, see Fig.~1 and Table 1.
(Various semi-empirical sources suggest nuclear matter saturation
to occur
at an energy per particle  ${\cal E}/A = -16 \pm 1$ MeV and a density
which is
equivalent to a Fermi momentum of $k_{F} = 1.35 \pm 0.05$ fm$^{-1}$
\cite{Bet71,Spr}.) This phenomenon is denoted
by the "Coester band" \cite{Coe}. The essential parameter of the
Coester band is the strength of the nuclear tensor force
as measured by the predicted D-state probability of the deuteron or
as expressed in terms of the wound integral in nuclear matter (see Table 1)
\cite{Mac89}.

\begin{figure}[t]
\centerline{\psfig{figure=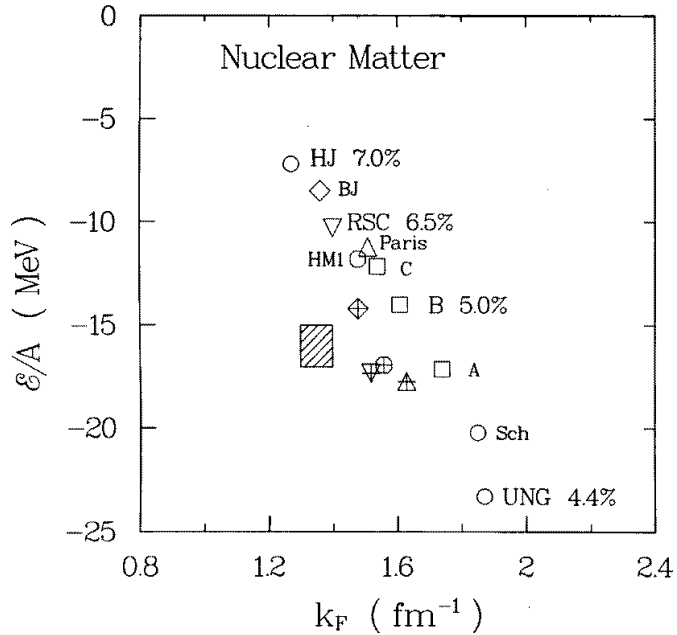,width=9cm}}
\caption{Nuclear matter saturation as predicted by a variety
of NN potentials (cf.\ Table 1). Open symbols are saturation
points obtained in the two-hole line approximation, 
symbols with a cross denote
corresponding predictions with three- and four-hole lines included.
\%-numbers refer to the D-state probability of the deuteron as
predicted by the corresponding potential. The shaded area denotes
the empirical value.}
\end{figure}

\begin{table}[t]
\centering
\footnotesize
\caption{Nuclear matter saturation properties as predicted
by various NN potentials.}
\begin{tabular}{llccccc}
\\  \hline\hline \\
 Potential & Ref.$^a$ & $P_{D}$ (\%) & $\kappa$ (\%) & ${\cal E}/A$ (MeV) & 
$k_{F}$ (fm$^{-1}$) & Ref.$^b$ 
 \\ \\
 \hline \\
HJ & 11 & 7.0 & 21 & --7.2 & 1.27 & 12\\
BJ & 13 & 6.6 & --- & --8.5 [--14.2] & 1.36 [1.48] & 12\\
RSC & 14 & 6.5 & 14 & --10.3$^{c}$ [--17.3] & 1.40$^{c}$ [1.52] & 15 [12]\\
V$_{14}$ & 16 & 6.1 & 12 [19] & --10.8 [--17.8] & 1.47 [1.62] & 17\\
Paris & 18 & 5.8 & 11 & --11.2 [--17.7] & 1.51 [1.63] & 18 [17]\\
HM1 & 19 & 5.8 & 11 & --11.8 [--16.9] & 1.48 [1.56] & 19 [17]\\
Sch & 20 & 4.9 & 8 & --20.2 & 1.85 & 19\\
UNG & 21 & 4.4 & 5 & --23.3 & 1.87 & 19\\
Bonn C & 10 & 5.6 & 8.1 & --12.1 & 1.54 & 10\\
Bonn B & 10 & 5.0 & 6.6 & --14.0 & 1.61 & 10\\
Bonn A & 10 & 4.4 & 5.4 & --17.1 & 1.74 & 10\\ 
\\ \hline\hline  \\
\multicolumn{6}{l}{\footnotesize
Given are the saturation energy per nucleon, ${\cal E}/A$, and Fermi 
momentum,} 
\\ \multicolumn{6}{l}{\footnotesize
$k_{F}$, as obtained in the two-hole line approximation using the standard} 
\\ \multicolumn{6}{l}{\footnotesize
choice for the single particle potential.  Results including three- and four-}
\\ \multicolumn{6}{l}{\footnotesize
hole line contributions are given in square brakets. The wound integral }
\\ \multicolumn{6}{l}{\footnotesize
$\kappa$ is given at $k_{F}=1.35$ fm$^{-1}$. $P_{D}$ is the predicted \%-D
state of the deuteron.}
\\ \multicolumn{6}{l}{\footnotesize
$^a$ References to the potentials.}
\\ \multicolumn{6}{l}{\footnotesize
$^b$ References for the nuclear matter calculations.}
\\ \multicolumn{6}{l}{\footnotesize
$^{c}$ Using OBEP for $J \geq 3$.}
\end{tabular}
\end{table}

The Brueckner-Goldstone expansion is believed to be convergent in terms
of the number of hole lines. Calculations by Day \cite{Day,Day81,Day85}
have confirmed this for the case of some realistic potentials.
However, three- and four-hole line diagrams contribute about
5-7 MeV to the binding energy per nucleon at saturation (cf. Table 1)
and, thus, are not negligible. In Fig. 1 open symbols represent
results obtained in the two-hole line approximation,
symbols with a cross denote results including the contributions
from three
and four hole-lines. It is seen that taking into account
up to four hole lines leads, indeed, to an improved Coester band
as compared to the two-hole line approximation;
however, the improvement is insufficient to explain the empirical
saturation point \cite{Day83}. Results based on the variational
approach are in fair agreement with Brueckner theory predictions
\cite{Day85} and, thus, also fail to quantitatively explain
nuclear saturation. 

Since the mid 1970's, there have been comprehensive efforts to check Brueckner 
theory \cite{DayJac}; we mention here, in particular, the work using 
hypernetted chain theory \cite{DayJac,Ben}. Based on this work, there are 
indications that the two hole-line approximation of Brueckner-Bethe theory may 
not necessarily be correct. Some published variational calculations are in 
fair agreement with Brueckner-Bethe results if three hole-lines plus a four 
hole-line estimate are included~\cite{Day85}.

Approaches discussed so far are based on the simplest model for the
atomic nucleus: Nucleons obeying the non-relativistic
Schr\"{o}dinger equation interact through a two-body potential
that fits low-energy NN scattering data and the properties of the
deuteron. The failure of this model to explain nuclear
saturation indicates that we may have to extend the model. One
possibility is to include degrees of freedom other than the
nucleon. The meson theory of the nuclear force suggests
to consider, particularly, meson and isobar degrees of freedom.
Characteristically, these degrees of freedom lead to medium
effects on the nuclear force when inserted into the
many-body problem as well as many-nucleon force contributions
(see Ref. \cite{Mac89} for a comprehensive review on this subject).
In general, the medium effects are repulsive whereas the many-nucleon
force contributions are attractive. Thus, there are
large cancellations and the net result is very small. The density
dependence of these effects/contributions is such that the saturation
properties of nuclear matter are not improved \cite{Mac89}.

We note that the discussion of many-body force effects in
the previous paragraph applies to approaches in which two- and many-body 
forces are treated on an equal footing; i.~e. both categories of forces
 are based on the same meson-baryon interactions and are treated 
consistently.
 The situation is different if the three-body force is introduced 
on an {\it ad hoc}  basis with the purpose to fit the empirical nuclear 
matter saturation~\cite{FP81}. Such a three-body potential may be large, 
particularly, if the two-body force used substantially underbinds nuclear 
matter. By construction such a three-body force improves the equation of state 
of nuclear matter as well as the description of light nuclei.

In the 1970's, a relativistic approach to
nuclear structure was developed by Miller and Green \cite{MG}.
They studied a Dirac-Hartree model for the groundstate of nuclei which was
able to reproduce the binding energies, the root-mean-square radii, and the
single-particle levels, particularly, the spin-orbit splittings. Their
potential consisted of a strong (attractive) scalar and (repulsive) vector
component. The Dirac-Hartree(-Fock) model was further developed by Brockmann
\cite{Bro} and by Horowitz and Serot \cite{HS81,SW}. At about that same time,
Clark and coworkers applied a Dirac equation containing a scalar and vector
field to proton-nucleus scattering \cite{Cla}. The most significant result of
this Dirac phenomenology is the quantitative fit of spin observables which were
only poorly described by the Schr\"{o}dinger equation \cite{Wal}.
 This success and
Walecka's theory on highly condensed matter \cite{Wal74} made relativistic
approaches very popular.

Inspired by these achievements,
a relativistic extension of Brueckner theory has been suggested by
Shakin
and collaborators \cite{Ana,CS}, frequently called the Dirac-Brueckner
approach. The advantage of a Brueckner theory is that the free
NN interaction is used; thus, there are no parameters in the force
which are
adjusted in the many-body problem. The essential point of the
Dirac-Brueckner approach is to use the Dirac equation for the
single particle motion in nuclear matter. In the work done by
the Brooklyn group the relativistic effect is calculated in first
order perturbation theory. This approximation is avoided and
a full self-consistency of the relativistic single-particle
energies and wave functions is performed in the subsequent work
 by Brockmann and Machleidt
\cite{BM,MB85,BM90}, and by ter Haar and Malfliet \cite{HM87}.
Formal aspects involved in the derivation of the relativistic G-matrix, have
been discussed in detail
by Horowitz and Serot \cite{HS84,HS87}.

The common feature of all Dirac-Brueckner results is that a
(repulsive) relativistic many-body effect is obtained which
is strongly density-dependent such that the empirical
nuclear matter saturation can be explained. In most calculations
a one-boson-exchange (OBE) potential is used
for the nuclear force. In Ref.~\cite{MB} a more realistic approach
to the NN interaction is taken applying  an explicit
model for the 2$\pi$-exchange that involves $\Delta$-isobars, thus,
avoiding the fictitious $\sigma$ boson typically used in OBE
models to provide intermediate range attraction. It is found in
Ref. \cite{MB} that the relativistic effect for the more realistic
model is almost exactly the same as that obtained for OBE potentials.
This finding seems to justify the use of the OBE model.

It is the purpose of this chapter to present a thorough introduction
into the Dirac Brueckner approach 
including the mathematical details of the
formalism involved.
Furthermore, we will present results for nuclear matter,
NN scattering in the nuclear medium, and finite nuclei.

\section{Relativistic Two-Nucleon Scattering}
\subsection{Covariant Equations}

Two-nucleon
scattering  is
described covariantly by
 the Bethe-Salpeter (BS) equation \cite{SB51}.
 In operator notation it may be written as
\begin{equation}
{\cal M=V+VGM}
\end{equation}
with ${\cal M}$ the invariant amplitude for the two-nucleon scattering process,
${\cal V}$ the sum of all connected two-particle irreducible diagrams and
${\cal G}$ the relativistic two-nucleon propagator. As this four-dimensional
integral equation is very difficult to solve \cite{FT75}, so-called
three-dimensional reductions have been proposed, which are more amenable to
numerical solution. Furthermore, it has been shown by Gross \cite{Gro82}
that the full BS equation in ladder approximation does not generate the
desired one-body equation in the one-body limit (i. e., when one of the
particles becomes very massive) in contrast to a large family of
three-dimensional quasi-potential equations.
 These approximations to the BS equation are also covariant
and satisfy relativistic elastic unitarity.
However, the three-dimensional
 reduction is not unique, and in principal infinitely many choices
exist \cite{Yae71}.
Typically they are derived by replacing Eq.~(1) by
 two coupled equations:
\begin{equation}
{\cal M=W+W}g{\cal M}
\end{equation}
and
\begin{equation}
{\cal W=V+V(G}-g){\cal W}
\end{equation}
where $g$ is a covariant three-dimensional propagator
with the same elastic unitarity cut as ${\cal G}$ in the physical region.
 In general,
the second term on the r.h.s.\ of Eq.\ (3) is dropped to arrive at a
real simplification of the  problem.

 Explicitly, we can write the BS equation for an arbitrary frame
(notation and conventions of Ref.~\cite{BD64})
\begin{equation}
\label{eq20}
{\cal M}(q',q|P)={\cal V}(q',q|P)+\int d^{4}k{\cal V}(q',k|P){\cal G}(k|P)
{\cal M}(k,q|P)
\end{equation}
with
\begin{eqnarray}
{\cal G}(k|P)&=&\frac{i}{(2\pi)^{4}}
\frac{1}{(\frac{1}{2}\not\!P+\not\!k-M+i\epsilon)^{(1)}}
\frac{1}{(\frac{1}{2}\not\!P-\not\!k-M+i\epsilon)^{(2)}}\\
             &=&\frac{i}{(2\pi)^{4}}
\left[\frac{\frac{1}{2}\not\!P+\not\!k+M}
{(\frac{1}{2}P+k)^{2}-M^{2}+i\epsilon}\right]^{(1)}
\left[\frac{\frac{1}{2}\not\!P-\not\!k+M}
{(\frac{1}{2}P-k)^{2}-M^{2}+i\epsilon}\right]^{(2)}
\end{eqnarray}
where $q$, $k$, and $q'$ are the initial, intermediate, and final relative
four-momenta, respectively (with e.\ g.\ $k=(k_{0},{\bf k})$),
 and $P=(P_0,{\bf P})$ is the total four-momentum;
$\not\!k=\gamma^{\mu}k_{\mu}$.
 The superscripts
refer to particle (1) and (2). In general,
we suppress spin (or helicity) and isospin indices.

 ${\cal G}$ and $g$ have the same discontinuity across the right hand cut,
if
\begin{eqnarray}
{\rm Im}\:{\cal G}(k|P)&=&-\frac{2\pi^2}{(2\pi)^{4}}
\left[\:\frac12\not\!P+\not\!k+M\:\right]^{(1)}\;
\left[\:\frac12\not\!P-\not\!k+M\:\right]^{(2)}\nonumber\\
&&\times
\delta^{(+)}\left[\:\left(\,\frac12\,P+k\right)^2-M^2\right]\;
\delta^{(+)}\left[\:\left(\,\frac12\,P-k\right)^2-M^2\right]\nonumber\\
&=&{\rm Im}\:g(k|P)
\end{eqnarray}
with $\delta^{(+)}$ indicating that only the positive energy root of the
argument of the $\delta$-function is to be included. From this follows easily
\begin{eqnarray}
{\rm Im}\:g(k|P)&=&-\frac1{8\pi^2}
\left[\:\frac12\not\!P+\not\!k+M\:\right]^{(1)}\;
\left[\:\frac12\not\!P-\not\!k+M\:\right]^{(2)}\nonumber\\
&&\times
\frac{\delta\left(\frac12P_0+k_0-E_{\frac12{\bf P}+{\bf k}}\right)
\delta\left(\frac12P_0-k_0-E_{\frac12{\bf P}-{\bf k}}\right)}
{4\,E_{\frac12{\bf P}+{\bf k}}\,E_{\frac12{\bf P}-{\bf k}}}
\end{eqnarray}
with $E_{\frac12{\bf P}\pm{\bf k}}\equiv\sqrt{M^2+(\frac12
{\bf P}\pm{\bf k})^2}$.
Using the equality
\begin{eqnarray}
\lefteqn{\hspace*{-.5in}
\delta\left(\frac12P_0+k_0-E_{\frac12{\bf P}+{\bf k}}\right)
\delta\left(\frac12P_0-k_0-E_{\frac12{\bf P}-{\bf k}}\right) = } \nonumber\\
&&\delta\left(P_0-E_{\frac12{\bf P}+{\bf k}}-
E_{\frac12{\bf P}-{\bf k}}\right)
\delta\left(k_0-\frac12 E_{\frac12{\bf P}+{\bf k}}+\frac12
E_{\frac12{\bf P}-{\bf k}}\right)
\end{eqnarray}
the imaginary part of the propagator $g(k|P)$ can now be written
\begin{eqnarray}
\label{eq25}
{\rm Im}\:g(k|P)&=&-\frac1{8\pi^2}
\frac{M^2}{E_{\frac12{\bf P}+{\bf k}}\: E_{\frac12{\bf P}-{\bf k}}}
\Lambda_+^{(1)}\left(\frac12{\bf P}+{\bf k}\right)\:
\Lambda_+^{(2)}\left(\frac12{\bf P}-{\bf k}\right)\nonumber\\
&&\times
\delta
\left(P_0-E_{\frac12{\bf P}+{\bf k}}-E_{\frac12{\bf P}-{\bf k}}\right)
\nonumber\\&&\times
\delta\left(k_0-\frac12 E_{\frac12{\bf P}+{\bf k}}+\frac12
E_{\frac12{\bf P}-{\bf k}}\right)
\end{eqnarray}
where
\begin{eqnarray}
\Lambda_+^{(i)}({\bf p})&=&\left(
\frac{\gamma^0E_{\bf p}-\mbox{\boldmath $\gamma$}\cdot{\bf p}+M}{2M}\right)^{(i)
}\\
&=&\sum_{\lambda_i}\:u({\bf p},\lambda_i)\overline u({\bf p},\lambda_i)
\end{eqnarray}
represents the positive-energy projection operator for nucleon $i$
($i=1,2$) with
$u({\bf p})$ a positive-energy Dirac spinor of momentum $\bf p$;
$\lambda_i$ denotes either the helicity or the spin projection of the
respective nucleon, and $E_{\bf p}=\sqrt{M^2+{\bf p}^2}$.
The projection operators imply that contributions involving virtual 
anti-nucleon
intermediate states (`pair terms') are suppressed. It has been shown 
in Refs.~\cite{FT80,ZT81} that these contributions are small when the 
pseudovector coupling is used for the pion.

 Note that
${\rm Im}\:g(k|P)$ is covariant, since
${\rm Im}\:g(k|P)={\rm Im}\:{\cal G}(k|P)$.

Using $\delta(P_0-E)=2E\delta(s-E^2+{\bf P}^2)$, where
$E=E_{\frac12{\bf P}+{\bf k}}+E_{\frac12{\bf P}-{\bf k}}$ and
$s=P^2=P_0^2-{\bf P}^2$, \ Eq.\ (10) can be re-written as
\begin{eqnarray}
{\rm Im}\:g(k|s)&=&-\frac{M^2}{8\pi^2}\,
\frac{2\left(E_{\frac12{\bf P}+{\bf k}}+E_{\frac12{\bf P}-{\bf k}}\right)}
{E_{\frac12{\bf P}+{\bf k}}E_{\frac12{\bf P}-{\bf k}}}
\Lambda_+^{(1)}\left(\frac12{\bf P}+{\bf k}\right)
\Lambda_+^{(2)}\left(\frac12{\bf P}-{\bf k}\right)\nonumber \\
&&\times
\delta\left[s-\left(E_{\frac12{\bf P}+{\bf k}}
+E_{\frac12{\bf P}-{\bf k}}\right)^2+{\bf P}^2\right]\nonumber \\
&&\times
\delta\left(k_0-\frac12 E_{\frac12{\bf P}+{\bf k}}+\frac12
E_{\frac12{\bf P}-{\bf k}}\right)
\end{eqnarray}

Knowing its imaginary part, we construct $g(k|s)$ by a
 dispersion integral
\begin{equation}
g(k|s)=\frac1{\pi}\int^{\infty}_{4M^2}\frac{ds'}{s'-s-i\epsilon}
{\rm Im}g(k|s')
\end{equation}

Inserting Eq.~(13) in Eq.~(14) and integrating, we obtain
\begin{eqnarray}
g(k|P)&=&-\frac{M^2}{(2\pi)^3}
\frac{2\left(E_{\frac12{\bf P}+{\bf k}}+E_{\frac12{\bf P}-{\bf k}}\right)}
{E_{\frac12{\bf P}+{\bf k}}E_{\frac12{\bf P}-{\bf k}}}\nonumber \\
&&\times\frac{\Lambda_+^{(1)}\left(\frac12{\bf P}+{\bf k}\right)
\Lambda_+^{(2)}\left(\frac12{\bf P}-{\bf k}\right)}{\left(
E_{\frac12{\bf P}+{\bf k}}+E_{\frac12{\bf P}-{\bf k}}\right)^2
-{\bf P}^2-s-i\epsilon}\nonumber \\
&&\times\delta(\begin{array}{c}
k_0-\frac12 E_{\frac12{\bf P}+{\bf k}}+\frac12
E_{\frac12{\bf P}-{\bf k}}
\end{array})
\end{eqnarray}
This three-dimensional propagator is known as the Blankenbecler-Sugar (BbS)
choice \cite{BS66,PL70}.
By construction, the propagator $g$ has the same
 discontinuity across the right-hand
cut as ${\cal G}$; therefore, it preserves the unitarity relation
satisfied by ${\cal M}$.

Using the
 angle averages $(\:\frac12{\bf P}\pm{\bf k}\:)^2\approx\frac14 {\bf P}^2 + 
{\bf k}^2$  and $(\:\frac12{\bf P}\pm{\bf q}\:)^2\approx\frac14 {\bf P}^2 + 
{\bf q}^2$,
 which should be a very good approximation,
Eq.~(15) assumes the much simpler form
\begin{equation}
\label{eqoben}
g(k|P)=\frac1{(2\pi)^3}\frac{M^2}{E_{\frac12{\bf P}+{\bf k}}}\frac{
\Lambda_+^{(1)}\left(\frac12{\bf P}+{\bf k}\right)
\Lambda_+^{(2)}\left(\frac12{\bf P}-{\bf k}\right)}
{E_{\frac12{\bf P}+{\bf q}}^2-E_{\frac12{\bf P}+{\bf k}}^2+i\epsilon}
\delta(k_0)
\end{equation}
where we used $s=4 E^2_{\frac12 {\bf P}+{\bf q}} - {\bf P}^2$.

Assuming ${\cal W}={\cal V}$,
the reduced Bethe-Salpeter equation, Eq.~(2), is obtained in explicit
form by replacing in Eq.~(4)
${\cal G}$ by $g$ of Eq.~(16),
yielding
\begin{eqnarray}
{\cal M}({\bf q'},{\bf q}|{\bf P})
&=&{\cal V}({\bf q'},{\bf q}|{\bf P})+
\int\frac{d^3k}{(2\pi)^3}{\cal V}({\bf q'},{\bf k}|{\bf P})
\frac{M^2}{E_{\frac12{\bf P}+{\bf k}}}
\nonumber \\
&&\times\frac{
\Lambda_+^{(1)}\left(\frac12{\bf P}+{\bf k}\right)
\Lambda_+^{(2)}\left(\frac12{\bf P}-{\bf k}\right)}
{E_{\frac12{\bf P}+{\bf q}}^2-E_{\frac12{\bf P}+{\bf k}}^2+i\epsilon}
{\cal M}({\bf k},{\bf q}|{\bf P})
\end{eqnarray}
in which both nucleons in intermediate states
are equally far off their mass shell.

Taking matrix elements between positive-energy spinors
 yields an equation for the scattering amplitude in an arbitrary frame
\begin{eqnarray}
{\cal T}({\bf q'},{\bf q}|{\bf P})
&=&V({\bf q'},{\bf q})+
\int\frac{d^3k}{(2\pi)^3}V({\bf q'},{\bf k})
\frac{M^2}{E_{\frac12{\bf P}+{\bf k}}}
\nonumber \\
&&\times\frac{1}
{E_{\frac12{\bf P}+{\bf q}}^2-E_{\frac12{\bf P}+{\bf k}}^2+i\epsilon}
{\cal T}({\bf k},{\bf q}|{\bf P})
\end{eqnarray}
where we used
\begin{eqnarray}
\lefteqn{\hspace*{-1.5in}
\begin{array}{l}
 \bar{u}_1(\frac12{\bf P}+{\bf q'})
\bar{u}_2(\frac12{\bf P}-{\bf q'})
{\cal V}({\bf q'},{\bf q}|{\bf P})
u_1(\frac12{\bf P}+{\bf q}) 
u_2(\frac12{\bf P}-{\bf q}) =
\end{array}
 } \nonumber\\
&& \bar{u}_1({\bf q'})
\bar{u}_2(-{\bf q'})
{\cal V}({\bf q'},{\bf q})
u_1({\bf q}) 
u_2(-{\bf q})\nonumber\\
&& \equiv V({\bf q'},{\bf q})
\end{eqnarray}
since this is a Lorentz scalar. An analogous statement applies to ${\cal T}$.
Calculations of nuclear matter and of finite nuclei
are performed in the rest frames of these systems. Thus, Eq.~(18) with the 
necessary medium modifications 
 would be appropriate for the evaluation of 
the nuclear matter reaction ($G$) matrix.

In the two-nucleon c.m. frame ( i.~e. for ${\bf P}=0$),
 the BbS propagator, Eq.~(16), reduces to
\begin{equation}
g({\bf k},s)=\frac{1}{(2\pi)^{3}}\frac{M^{2}}{E_{\bf k}}
\frac{\Lambda_{+}^{(1)}({\bf k})
\Lambda_{+}^{(2)}({\bf -k})}
{\frac{1}{4}s-E^{2}_{\bf k}+i\epsilon}\delta{(k_0)}
\end{equation}
which implies the scattering equation
\begin{equation}
\label{bbs}
{\cal T}({\bf q'},{\bf q})=V({\bf q'},{\bf q})+
\int \frac{d^{3}k}{(2\pi)^{3}}
V({\bf q'},{\bf k})
\frac{M^{2}}{E_{\bf k}}
\frac{1}
{{\bf q}^{2}-{\bf k}^{2}+i\epsilon}
{\cal T}({\bf k},{\bf q})
\end{equation}
Two-nucleon scattering is considered most conveniently in the two-nucleon c.m. 
frame; thus, for calculations of free space 
two-nucleon scattering in the BbS approximation,
 one would  use Eq.~(21).

Defining
\begin{equation}
\hat{T}({\bf q'}, {\bf q})
 = \sqrt{\frac{M}{E_{q'}}}\: {\cal T}({\bf q'}, {\bf q})\:
 \sqrt{\frac{M}{E_{q}}}
\label{21.16}
\end{equation}
and
\begin{equation} 
\hat{V}({\bf q'},{\bf q})
 = \sqrt{\frac{M}{E_{q'}}}\:  V({\bf q'},{\bf q})\:
 \sqrt{\frac{M}{E_{q}}},
\label{21.17}
\end{equation}
which has become known as ``minimal relativity''~\cite{BJK69},
we can rewrite Eq.~(21) as
\begin{equation}
\hat{T}({\bf q'},{\bf q})=\hat{V}({\bf q'},{\bf q})+
\int d^3k\:
\hat{V}({\bf q'},{\bf k})\:
\frac{M}
{{\bf q}^{2}-{\bf k}^{2}+i\epsilon}\:
\hat{T}({\bf k},{\bf q})
\label{21.18}
\end{equation}
which has the form of the non-relativistic Lippmann-Schwinger
equation.
A potential defined within an equation which is formally
identical to the (non-relativistic) Lippmann-Schwinger equation,
can then be applied in conventional (non-relativistic) nuclear structure
physics.
 This  is the practical relevance of Eqs.~(22-24).
On the other hand, for relativistic nuclear structure physics, Eq.~(21)
is the starting point.

The BbS  propagator
is the most widely used approximation.
Another choice, that has been frequently applied,
is the version suggested by Thompson \cite{Tho70}.
The manifestly covariant form of Thompson's propagator is the same
as Eq.~(14), but with $\int^{\infty}_{4M^2}ds'/(s'-s-i\epsilon)$
 replaced by
$\int^{\infty}_{2M}d\sqrt{s'}/(\sqrt{s'}-\sqrt{s}-i\epsilon)$.
 After taking the angle average for
$|\:\frac12{\bf P}\pm
{\bf k}\:|$ and $|\:\frac12{\bf P}\pm
{\bf q}\:|$, this propagator reads 
\begin{equation}
g(k|P)=\frac1{(2\pi)^3}\frac{M^2}{E_{\bf k}E_{\frac12{\bf P}+{\bf k}}}
\frac{
\Lambda_+^{(1)}\left(\frac12{\bf P}+{\bf k}\right)
\Lambda_+^{(2)}\left(\frac12{\bf P}-{\bf k}\right)}{2E_{\bf q}-2E_{\bf k}
+i\epsilon}
\delta(k_0)
\end{equation}
The equation for the scattering amplitude in an arbitrary frame is
\begin{eqnarray}
\label{thompson}
{\cal T}({\bf q'},{\bf q}|{\bf P})
&=& V ({\bf q'},{\bf q})+
\int\frac{d^3k}{(2\pi)^3} V ({\bf q'},{\bf k})
\frac{M^2}{E_{\bf k}\:E_{\frac12{\bf P}+{\bf k}}}\nonumber\\
&&\times\frac1{2E_{\bf q}-2E_{\bf k}
+i\epsilon}                         
{\cal T}({\bf k},{\bf q}|{\bf P})
\end{eqnarray}
For calculations in the rest frame of nuclear matter or finite nuclei, this
equation together with the necessary medium modifications
($M\longmapsto\tilde{M}$, Pauli projector $Q$)
 is appropriate (see 
Eq.~(95) below). 
In our actual calculations in nuclear matter, we replace 
$E_{\bf k}$ by $E_{\frac12{\bf P}+{\bf k}}$ and
$E_{\bf q}$ by $E_{\frac12{\bf P}+{\bf q}}$ in the denominator of Eq.~(26). 
This replacement makes possible an interpretation of the energy 
denominator in terms of differences between single particle energies
which are typically defined in the rest frame of the many-body system.
 This 
allows for a consistent application of this equation in nuclear matter {\it 
and} finite nuclei~\cite{MMB90}. 
The change of the numerical results by this replacement
is negligibly small (less than 0.1 MeV for the energy per nucleon at nuclear 
matter denisty),
 since the factor 
$\frac{M^2}{E_{\bf k}\:E_{\frac12{\bf P}+{\bf k}}}$
is slightly reduced, while the term
$(2E_{\bf q}-2E_{\bf k}+i\epsilon)^{-1}$ is slightly enhanced.
                
In the two-nucleon c.m. frame (${\bf P}=0$), the Thompson equation is
\begin{equation}
{\cal T}({\bf q'},{\bf q})=V({\bf q'},{\bf q})+
\int \frac{d^{3}k}{(2\pi)^{3}}
V({\bf q'},{\bf k})
\frac{M^{2}}{E_{\bf k}^2}
\frac{1}
{2E_{\bf q}-2E_{\bf k}+i\epsilon}
{\cal T}({\bf k},{\bf q}),
\end{equation}
This equation is usefull for free-space two-nucleon scattering. In the 
calculations of this chapter, always Thompson's equations are used.

Again, we may introduce some special definitions. In the case of the Thompson 
equation, it is convenient to define
\begin{equation}
\check{T}({\bf q'}, {\bf q})
 = {\frac{M}{E_{q'}}}\: {\cal T}({\bf q'}, {\bf q})\:
 {\frac{M}{E_{q}}}
\label{21.22}
\end{equation}
and
\begin{equation} 
\check{V}({\bf q'},{\bf q})
 = {\frac{M}{E_{q'}}}\:  V({\bf q'},{\bf q})\:
 {\frac{M}{E_{q}}}\: .
\label{21.23}
\end{equation}
With this,
we can rewrite Eq.~(27) as
\begin{equation}
\check{T}({\bf q'},{\bf q})=\check{V}({\bf q'},{\bf q})+
\int d^3k\:
\check{V}({\bf q'},{\bf k})
\frac{1}
{2E_q-2E_k+i\epsilon}
\check{T}({\bf k},{\bf q})
\label{21.24}
\end{equation}
which looks like a Lippmann-Schwinger equation with 
relativistic energies in the 
propagator.

Since both nucleons are equally off-shell in
 the BbS or Thompson equation, the exchanged bosons tranfer
 three-momentum only, i.\ e.\
the meson propagator is
(for a scalar exchange)
\begin{equation}
\frac{i}{-({\bf q'-q})^{2}-m_{\alpha}^{2}};
\end{equation}
this is also referred to as a static (or non-retarded) propagator.

Many more choices for $g$ have been suggested in the literature.
For most of the other choices, the three-dimensional two-nucleon
propagator is either of the BbS or the Thompson form.
However, there may be differences in the $\delta$ function
in Eqs.~(20) and (25). For example, Schierholz \cite{Sch} and
Erkelenz \cite{Erk74} propose $\delta(k_{0}+\frac{1}{2}\sqrt{s}
-E_{k})$, restricting one particle to its mass shell.  It implies
the meson propagator
\begin{equation}
\frac{i}{(E_{{\bf q}'}-E_{{\bf q}})^{2} -({\bf q'-q})^{2}-m_{\alpha}^{2}}.
\end{equation}
However, it has been shown \cite{Mac82} (see also Appendix E.1 of
Ref. \cite{MHE87})
 that the term $(E_{{\bf q}'}-E_{{\bf q}})^{2}$
in this propagator has an effect which is opposite to the one obtained when
treating meson retardation properly.
Also, the medium effect on meson propagation in nuclear matter
which, when calculated correctly, is repulsive \cite{Foot1}, comes out
attractive when using the form Eq.~(32) \cite{Foot2}.
Thus, in spite of its suggestive
appearance and in spite of early believes, the meson propagator
Eq.~(32) has nothing to do with genuine meson retardation and,
therefore, should be discarded.
This remark applies to the scattering of two particles of equal mass.
If one particle is much heavier than the other one,
 it may, however, be appropriate to put one particle
(namely, the more massive one) on its mass shell.

A thorough discussion of the Bethe-Salpeter equation and/or
a systematic study of a large family of possible relativistic
three-di\-men\-sional reductions can be found in
Refs. \cite{Nak69,WJ73,BJ76}.
Tjon and coworkers have compared results obtained by solving the full
four-dimensional Bethe-Salpeter equation applying a full set of OBE diagrams
with those from the BbS and some other three-dimensional
 equations; for BbS they find only small
differences as compared to full BS \cite{FT80,ZT81}.
This is also true for the Thompson choice, since it differs little
from BbS.

Notice that the quantity $\cal T$ in Eqs.~(21) and (27)
 is invariant, while 
$\hat{T}$ of Eq.~(24) and $\check{T}$ of Eq.~(30) are not.
 $\hat{T}$ is equivalent to the
familiar  non-relativistic $T$-matrix.

The relationship of the invariant $\cal T$ to the $S$-matrix is
\begin{eqnarray}
\langle p_{1}' p_{2}' | S | p_{1} p_{2} \rangle  = 
\langle p_{1}' p_{2}' |  p_{1} p_{2} \rangle 
-2\pi i \delta^{(4)}(p_{1}'+p_{2}'-p_{1}-p_{2})
\frac{M^{2}}{E_{q}^{2}}{\cal T}({\bf q',q})
\nonumber \\
\mbox{} \label{21.25}
\end{eqnarray}
with $p_{i}$ 
the initial and $p_{i}'$ ($i=1,2$) the final four momenta of the two
interacting nucleons. 
The normalization is $\langle p'|p\rangle = \delta^{(3)}({\bf p'-p})$.

\subsection{R-Matrix Equation and Helicity State Basis}

On a computer, real analysis is much faster than complex analysis.
It is therefore desirable to deal with real quantities whenever possible. 
The scattering amplitude below particle production threshold can be expressed 
in terms of the real $R$-matrix (better known as the `$K$-matrix') 
which is defined by
\begin{equation}
T=R-i\pi R\delta(E-H_0)T
\label{22.0}
\end{equation}
The equation for the real $\hat{R}$-matrix corresponding to the complex 
$\hat{T}$-matrix of Eq.~(24) is
\begin{equation}
\hat{R}({\bf q'},{\bf q})=\hat{V}({\bf q'},{\bf q})+
{\cal P} \int d^3k\:
\hat{V}({\bf q'},{\bf k})\:
\frac{M}
{{\bf q}^{2}-{\bf k}^{2}}\:
\hat{R}({\bf k},{\bf q})
\label{22.00}
\end{equation}
where ${\cal P}$ denotes the principal value.

Now, we also need to consider the spin of the nucleons explicitly.
The easiest way to treat the spin-projections of spin-$\frac12$ particles in 
a covariant way is to use the helicity representation.
In our further developments, we will therefore use a helicity state basis
(cf.\ Appendix C of 
Ref.~\cite{MHE87} which is based upon Refs.~\cite{JW59,EAH71}.

The helicity $\lambda_i$ of particle $i$ (with $i=1$ or 2) 
is the eigenvalue of the 
helicity operator $\frac12 {\bf \sigma}_i \cdot {\bf p}_i/|{\bf p}_i|$ which is 
$\pm \frac12$. 

Using helicity states, the $\hat{R}$-matrix equation reads,
 after partial wave decomposition,
\begin{eqnarray}
\langle \lambda_1' \lambda_2'|\hat{R}^J(q',q)|\lambda_1 \lambda_2 \rangle
&=&\langle \lambda_1' \lambda_2'|\hat{V}^J(q',q)|\lambda_1 \lambda_2 \rangle
\nonumber \\
 & & \mbox{} + \sum_{h_1,h_2} {\cal P} \int^{\infty}_0 dk\: k^2
 \frac{M}{q^2-k^2}
\langle \lambda_1' \lambda_2'|\hat{V}^J(q',k)|h_1 h_2 \rangle \nonumber \\
 & & \mbox{} \times
\langle h_1 h_2|\hat{R}^J(k,q)|\lambda_1 \lambda_2 \rangle 
\label{22.1}
\end{eqnarray}
where $J$ denotes the total angular momentum of the two nucleons.
Here and throughout the rest of this chapter,
momenta denoted by non-bold letters are the magnitudes of
three-momenta, e.~g. 
$q\equiv|{\bf q}|$, $k\equiv|{\bf k}|$, etc.;
$h_1$ and $h_2$ are the helicities in intermediate states for nucleon 1 and 2, 
respectively.
Equation~(\ref{22.1}) is a system 
of coupled integral equations which needs to be solved to obtain the desired
matrix elements of $\hat{R}^J$.

Ignoring anti-particles, there are $4\times 4 = 16$
 helicity amplitudes for $\hat{R}^J$. However,
time-reversal invariance, parity conservation, and the fact that we are 
dealing with two identical fermions imply that 
only six amplitudes are independent. For these six amplitudes, we 
choose the following set:
\begin{eqnarray}
\hat{R}^J_1(q',q)&\equiv& \langle ++|\hat{R}^J(q',q)|++ \rangle \nonumber \\
\hat{R}^J_2(q',q)&\equiv& \langle ++|\hat{R}^J(q',q)|-- \rangle \nonumber \\
\hat{R}^J_3(q',q)&\equiv& \langle +-|\hat{R}^J(q',q)|+- \rangle \nonumber \\
\hat{R}^J_4(q',q)&\equiv& \langle +-|\hat{R}^J(q',q)|-+ \rangle  \\
\hat{R}^J_5(q',q)&\equiv& \langle ++|\hat{R}^J(q',q)|+- \rangle \nonumber \\
\hat{R}^J_6(q',q)&\equiv& \langle +-|\hat{R}^J(q',q)|++ \rangle \nonumber
\label{22.2}
\end{eqnarray}
where $\pm$ stands for $\pm \frac12$.
Notice that
\begin{equation}
\hat{R}^J_5(q',q)=\hat{R}^J_6(q,q') .
\label{22.3}
\end{equation}

We have now six coupled equations.
To partially decouple this system, it is usefull to 
introduce the following linear combinations of helicity amplitudes:
\begin{eqnarray}
^0\hat{R}^J&\equiv&\hat{R}^J_1 - \hat{R}^J_2  \nonumber \\
^1\hat{R}^J&\equiv&\hat{R}^J_3 - \hat{R}^J_4  \nonumber \\
^{12}\hat{R}^J&\equiv&\hat{R}^J_1 + \hat{R}^J_2 \label{22.4} \\
^{34}\hat{R}^J&\equiv&\hat{R}^J_3 + \hat{R}^J_4 \nonumber \\
^{55}\hat{R}^J&\equiv&2\hat{R}^J_5 \nonumber \\
^{66}\hat{R}^J&\equiv&2\hat{R}^J_6 \nonumber 
\end{eqnarray}
We also introduce corresponding definitions for $\hat{V}^J$.
(Of course, analogous definitions exist for $R, \check{R}, V,
\mbox{ and } \check{V}$.)                   
Using these definitions, Eq.~(\ref{22.1}) decouples into the following 
three sub-systems of integral equations:\\
{ Spin singlet}
\begin{equation}
^0\hat{R}^J(q',q) =\:  ^0\hat{V}^J(q',q)+{\cal P}\int^{\infty}_0 dk\: k^2
\frac{M}{q^2-k^2}\: ^0\hat{V}^J(q',k)\: ^0\hat{R}^J(k,q)
\label{22.5} 
\end{equation}
{ Uncoupled spin triplet}
\begin{equation}
^1\hat{R}^J(q',q) =\:  ^1\hat{V}^J(q',q)+{\cal P}\int^{\infty}_0 dk\: k^2
\frac{M}{q^2-k^2}\: ^1\hat{V}^J(q',k)\: ^1\hat{R}^J(k,q)
\label{22.6} 
\end{equation}
{ Coupled triplet states}
\begin{eqnarray}
^{12}\hat{R}^J(q',q)&=& ^{12}\hat{V}^J(q',q)+{\cal P}\int^{\infty}_0 dk\: k^2
\frac{M}{q^2-k^2}[\: ^{12}\hat{V}^J(q',k)\: ^{12}\hat{R}^J(k,q) \nonumber \\
 & & \mbox{} +\: ^{55}\hat{V}^J(q',k)\: ^{66}\hat{R}^J(k,q)]  \nonumber \\
^{34}\hat{R}^J(q',q)&=& ^{34}\hat{V}^J(q',q)+{\cal P}\int^{\infty}_0 dk\: k^2
\frac{M}{q^2-k^2}[\: ^{34}\hat{V}^J(q',k)\: ^{34}\hat{R}^J(k,q) \nonumber \\
 & & \mbox{} +\: ^{66}\hat{V}^J(q',k)\: ^{55}\hat{R}^J(k,q)]   \nonumber \\
^{55}\hat{R}^J(q',q)&=& ^{55}\hat{V}^J(q',q)+{\cal P}\int^{\infty}_0 dk\: k^2
\frac{M}{q^2-k^2}[\: ^{12}\hat{V}^J(q',k)\: ^{55}\hat{R}^J(k,q) \nonumber \\
 & & \mbox{} +\: ^{55}\hat{V}^J(q',k)\: ^{34}\hat{R}^J(k,q)]  \nonumber \\
^{66}\hat{R}^J(q',q)&=& ^{66}\hat{V}^J(q',q)+{\cal P}\int^{\infty}_0 dk\: k^2
\frac{M}{q^2-k^2}[\: ^{34}\hat{V}^J(q',k)\: ^{66}\hat{R}^J(k,q) \nonumber \\
 & & \mbox{} +\: ^{66}\hat{V}^J(q',k)\: ^{12}\hat{R}^J(k,q)]  \nonumber \\
\mbox{} \label{22.7}
\end{eqnarray}

More common in nuclear physics is the representation of 
two-nucleon states
in terms of an 
$|LSJM\rangle$ basis, 
where $S$ denotes the total spin, $L$ the total orbital 
angular momentum, and $J$ the total angular momentum with 
projection $M$. 
In this basis, we will denote the $\hat{R}$ matrix elements by
$\hat{R}^{JS}_{L',L}\equiv \langle L'SJM|\hat{R}|LSJM\rangle$.
  These are obtained from the helicity state matrix 
elements by the following unitary transformation:\\
{ Spin singlet}
\begin{equation}
\hat{R}^{J0}_{J,J}\: =\: ^0\hat{R}^J
\label{22.10}
\end{equation}
{ Uncoupled spin triplet} 
\label{22.11}
\begin{equation}
\hat{R}^{J1}_{J,J}\: =\: ^1\hat{R}^J
\end{equation}
{ Coupled triplet states}
\begin{eqnarray}
\hat{R}^{J1}_{J-1,J-1} & = & \frac{1}{2J+1} \left[J\: ^{12}\hat{R}^J 
+ (J+1)\: ^{34}\hat{R}^J
+ \sqrt{J(J+1)}(\: ^{55}\hat{R}^J+\: ^{66}\hat{R}^J)\right]
\nonumber \\
\hat{R}^{J1}_{J+1,J+1} & = & \frac{1}{2J+1} \left[(J+1)\: ^{12}\hat{R}^J 
+ J\: ^{34}\hat{R}^J
- \sqrt{J(J+1)}(\: ^{55}\hat{R}^J+\: ^{66}\hat{R}^J)\right]
\nonumber \\
\hat{R}^{J1}_{J-1,J+1} & = & \frac{1}{2J+1} \left[\sqrt{J(J+1)}
(\: ^{12}\hat{R}^J -\: ^{34}\hat{R}^J)
- J\: ^{55}\hat{R}^J + (J+1)\: ^{66}\hat{R}^J)\right]
\nonumber  \\
\hat{R}^{J1}_{J+1,J-1} & = & \frac{1}{2J+1} \left[\sqrt{J(J+1)}
(\: ^{12}\hat{R}^J -\: ^{34}\hat{R}^J)
+ (J+1)\: ^{55}\hat{R}^J - J\: ^{66}\hat{R}^J)\right]
\nonumber \\
\mbox{} \label{22.12} 
\end{eqnarray}
Analogous transformations exist for $\hat{V}, R, \check{R}, V,
\mbox{ and } \check{V}$.                   

Instead of solving the coupled system in the form 
Eq.~(\ref{22.7}), one can also first 
apply the transformation Eq.~(\ref{22.12}) to $\hat{V}$ and $\hat{R}$
in Eq.~(\ref{22.7})  
yielding a  system of four coupled integral equations for
$\hat{R}$  in $LSJ$ representation,
\begin{eqnarray}
\hat{R}^{J1}_{++}(q',q)&=& \hat{V}^{J1}_{++}(q',q)
+{\cal P}\int^{\infty}_0 dk\: k^2\: \frac{M}{q^2-k^2}
[\: \hat{V}^{J1}_{++}(q',k)\: \hat{R}^{J1}_{++}(k,q)  
\nonumber \\  & & \mbox{} 
+\: \hat{V}^{J1}_{+-}(q',k)\: \hat{R}^{J1}_{-+}(k,q)]  
\nonumber \\
\hat{R}^{J1}_{--}(q',q)&=& \hat{V}^{J1}_{--}(q',q)
+{\cal P}\int^{\infty}_0 dk\: k^2\: \frac{M}{q^2-k^2}
[\: \hat{V}^{J1}_{--}(q',k)\: \hat{R}^{J1}_{--}(k,q) 
\nonumber \\  & & \mbox{}
+\: \hat{V}^{J1}_{-+}(q',k)\: \hat{R}^{J1}_{+-}(k,q)]   
\nonumber \\
\hat{R}^{J1}_{+-}(q',q)&=& \hat{V}^{J1}_{+-}(q',q)
+{\cal P}\int^{\infty}_0 dk\: k^2\: \frac{M}{q^2-k^2}
[\: \hat{V}^{J1}_{++}(q',k)\: \hat{R}^{J1}_{+-}(k,q) 
\nonumber \\ & & \mbox{} 
+\: \hat{V}^{J1}_{+-}(q',k)\: \hat{R}^{J1}_{--}(k,q)]  
\nonumber \\
\hat{R}^{J1}_{-+}(q',q)&=& \hat{V}^{J1}_{-+}(q',q)
+{\cal P}\int^{\infty}_0 dk\: k^2\: \frac{M}{q^2-k^2}
[\: \hat{V}^{J1}_{--}(q',k)\: \hat{R}^{J1}_{-+}(k,q) 
\nonumber \\ & & \mbox{} 
+\: \hat{V}^{J1}_{-+}(q',k)\: \hat{R}^{J1}_{++}(k,q)]  
\nonumber \\
\mbox{} \label{22.13}
\end{eqnarray}
where we used the abbreviations \\ 
$\hat{R}^{J1}_{++}\equiv \hat{R}^{J1}_{J+1,J+1},\;
\hat{R}^{J1}_{--}\equiv \hat{R}^{J1}_{J-1,J-1},\;
\hat{R}^{J1}_{+-}\equiv \hat{R}^{J1}_{J+1,J-1},\;
\hat{R}^{J1}_{-+}\equiv \hat{R}^{J1}_{J-1,J+1}$.\\
Conventionally, the coupled triplet channels
in NN scattering are considered in this form. 
It can also be obtained by decomposing Eq.~(\ref{22.00}) directly into $LSJ$ 
states. In a non-relativistic consideration, it is the tensor 
force which couples triplet states with $L=J\pm1$.

So far, we have never mentioned the total isospin of the two-nucleon system,
$T$ (which is either 0 or 1). The 
reason for this is simply that $T$ is not an independent quantum number.
 Namely, due to the 
antisymmetry of the two-fermion state, the quantum numbers $L, S \mbox{ and }
T$ have to fulfill the condition
\begin{equation}
(-1)^{L+S+T}=-1\: .
\label{22.14}
\end{equation}
Thus, for given $L$ and $S$, $T$ is fixed.

\paragraph*{On-Shell R-Matrix and Phase Shifts.}
Phase shifts are a parametrization of the unitary $S$-matrix which for 
uncoupled cases is given by
\begin{equation}
S_J = e^{2i\delta_J} .
\label{23.0}
\end{equation}
Using the above and Eqs.~(\ref{21.25}) 
and (\ref{22.0}) in partial wave decomposition, 
one can relate the on-energy-shell
$R$-matrix to the phase shifts
as follows:\\
{ Spin singlet}
\begin{equation}
\tan\: ^0\delta^J(E_{lab}) = -\frac{\pi}{2} q M\: ^0\hat{R}^J(q,q)
\label{23.1}
\end{equation}
{ Uncoupled spin triplet}
\begin{equation}
\tan\: ^1\delta^J(E_{lab}) = -\frac{\pi}{2} q M\: ^1\hat{R}^J(q,q)
\label{23.2}
\end{equation}
For the {\it coupled states}, a unitary transformation is needed to diagonalize 
the two-by-two coupled $R$-matrix. This requires an additional parameter,
known as the `mixing 
parameter' $\epsilon_J$. Using the convention introduced by Blatt and 
Biedenharn~\cite{BB52}, the eigenphases for the coupled channels are,
in terms of the on-shell $\hat{R}$-matrix,
\begin{eqnarray}
\tan\: \delta^J_{\mp}(E_{lab})& =& -\frac{\pi}{4} q M
\left[ \hat{R}^J_{--} + \hat{R}^J_{++} \pm
\frac{\hat{R}^J_{--} - \hat{R}^J_{++}}{\cos 2\epsilon_J}\right]
\nonumber \\
\label{23.3}  \\
\tan 2\epsilon_J(E_{lab}) & = & \frac{2\hat{R}^J_{+-}}
{\hat{R}^J_{--} - \hat{R}^J_{++}}\: . \nonumber
\end{eqnarray}
All $\hat{R}$-matrix elements in these formulae carry the arguments
 $(q,q)$ where $q$ denotes the c.m.~on-energy-shell momentum that
 is related to the energy in the 
laboratory system, $E_{lab}$, by
\begin{equation}
E_{lab} = {2q^2}/{M} .
\label{23.4}
\end{equation}  
An alternative convention for the phase parameters has been 
introduced by Stapp {\it 
et al.}~\cite{SYM57}, known as `bar' phase shifts. These are related to 
the Blatt-Biedenharn parameters by
\begin{eqnarray}
\bar{\delta}^J_+ + \bar{\delta}^J_- & = & \delta^J_+ + \delta^J_- 
\nonumber \\
\sin (\bar{\delta}^J_- - \bar{\delta}^J_+) & = & 
{\tan 2\bar{\epsilon}_J}/{\tan 2\epsilon_J}
\label{23.5} \\
\sin (\delta^J_- - \delta^J_+) & = & 
{\sin 2\bar{\epsilon}_J}/{\sin 2\epsilon_J}
\nonumber
\end{eqnarray}
We use the `bar' convention.

Using the transformation Eq.~(\ref{22.12}), the phase parameters for the 
coupled case can also be expressed directly 
in terms of the helicity-state $\hat{R}$-matrix 
elements; one obtains
\begin{eqnarray}
\tan\: \delta^J_{\mp}(E_{lab})& =& -\frac{\pi}{4} q M
\left[\: ^{12}\hat{R}^J +\: ^{34}\hat{R}^J \mp
\frac{^{12}\hat{R}^J -\: ^{34}\hat{R}^J - 4\sqrt{J(J+1)}\: 
^{55}\hat{R}^J}{(2J+1)\cos 2\epsilon_J}\right]
\nonumber \\
\label{23.6} \\
\tan 2\epsilon_J(E_{lab}) & = & -2\frac{\sqrt{J(J+1)}(\: 
^{12}\hat{R}^J -\: ^{34}\hat{R}^J) +\: 
^{55}\hat{R}^J}{^{12}\hat{R}^J -\: 
^{34}\hat{R}^J - 4\sqrt{J(J+1)}\: ^{55}\hat{R}^J}\; .
\nonumber
\end{eqnarray}

\paragraph*{Effective Range Parameters.}
For low-energy $S$-wave scattering, $q\cot \delta$ can be expanded as a 
function of $q$
\begin{equation}
\frac{q}{\tan \delta}=q\cot \delta\approx -\frac{1}{a}+\frac12 r q^2
\label{24.1}
\end{equation}
where $a$ is called the scattering length and $r$ the effective range.

Rewriting this equation for two different 
(small) on-shell momenta $q_1$ and $q_2$
(with $E^{(i)}_{lab} = {2q_i^2}/{M}$ and phase shifts $\delta_i$, $i=1,2$),
we can determine the two unknown constants $a$ and $r$,
\begin{equation}
r = \frac{4}{M}\: \frac{\frac{q_1}{\tan \delta_1}-\frac{q_2}{\tan \delta_2}}
{E^{(1)}_{lab}-E^{(2)}_{lab}}
\label{24.2}
\end{equation}
and
\begin{equation}
\frac{1}{a} = \frac{M}{4}\: r\: E^{(i)}_{lab} - \frac{q_i}{\tan \delta_i}
\label{24.3}
\end{equation}
with $i=1$ or 2.

\paragraph*{Using Thompson's Equation.}
The Thompson equation, Eq.~(27),
 is solved most conveniently in `check' notation, 
Eqs.~(\ref{21.22})-(\ref{21.24}). The corresponding
equations for the $\check{R}$-matrix are 
obtained from Eqs.~(\ref{22.5})-(\ref{22.7}) by replacing
\begin{equation}
\frac{M}{q^2-k^2}  \longmapsto  \frac{1}{2E_q-2E_k} = 
\frac{\frac12 (E_q+E_k)}{q^2-k^2} 
\label{25.1} 
\end{equation}
and
\begin{equation}
\hat{R}     \longmapsto  \check{R}
\label{25.0}
\end{equation}
This is easily understood by comparing Eq.~(\ref{21.18}) with 
Eq.~(\ref{21.24}). $\check{R}$ is defined in analogy to Eq.~(\ref{21.22}).
\\
The phase shift relation is
\begin{equation}
\tan\: ^0\delta^J(E_{lab}) = -\frac{\pi}{2} q E_q\: ^0\check{R}^J(q,q)
\label{25.2}
\end{equation}
and similarly for the other channels.

\section{One-Boson-Exchange Potentials}

\subsection{Interaction Lagrangians and OBE Amplitudes}

We use the following Lagrangians for meson-nucleon coupling
\begin{eqnarray}
{\cal L}_{pv}&=& -\frac{f_{ps}}{m_{ps}}\bar{\psi}
\gamma^{5}\gamma^{\mu}\psi\partial_{\mu}\varphi^{(ps)}
\label{31.2}\\
{\cal L}_{s}&=& +g_{s}\bar{\psi}\psi\varphi^{(s)}
\label{31.3}\\
{\cal L}_{v}&=&-g_{v}\bar{\psi}\gamma^{\mu}\psi\varphi^{(v)}_{\mu}
-\frac{f_{v}}{4M} \bar{\psi}\sigma^{\mu\nu}\psi(\partial_{\mu}
\varphi_{\nu}^{(v)}
-\partial_{\nu}\varphi_{\mu}^{(v)})
\label{31.4}
\end{eqnarray}
with $\psi$ the nucleon and $\varphi^{(\alpha)}_{(\mu)}$ the meson fields
(notation and conventions as in Ref.~\cite{BD64}).
For isospin 1 mesons, $\varphi^{(\alpha)}$ is to be replaced
by {\boldmath $\tau \cdot \varphi^{(\alpha)}$},
with $\tau^{l}$ ($l=1,2,3$) the usual Pauli matrices.
$ps$, $pv$, $s$, and $v$ denote pseudoscalar, pseudovector, scalar, and vector
coupling/field, respectively.

The one-boson-exchange potential (OBEP) is defined as a sum of
one-particle-exchange amplitudes of certain bosons
with given mass and coupling. We use the six non-strange bosons with masses
below 1 GeV/c$^2$. Thus,
\begin{equation}
V_{OBEP}=\sum_{\alpha=\pi,\eta,\rho,\omega,\delta,\sigma}
V^{OBE}_{\alpha}
\label{31.5}
\end{equation}
with $\pi$ and $\eta$ pseudoscalar,
$\sigma$ and $\delta$ scalar, and
$\rho$ and $\omega$ vector particles.
The contributions from the iso-vector bosons $\pi, \delta$ and $\rho$
contain a factor {\boldmath $\tau_{1} \cdot \tau_{2}$}.

The above Lagrangians imply the following 
 OBE amplitudes:\footnote{Strictly speaking, we give here the potential 
defined as $i$ times the Feynman amplitude; furthermore, there is a factor of 
$i$ for each vertex and propagator; since $i^4=1$, we simply ignore these 
factors of $i$.}
\footnotesize
\begin{eqnarray}
\lefteqn{\hspace{-.6cm}\langle {\bf q'} \lambda_{1}'\lambda_{2}'|V^{OBE}_{pv}|
{\bf q}\lambda_{1}\lambda_{2}\rangle} \nonumber \\
 & = &
\frac{1}{(2\pi)^3} \frac{f^{2}_{ps}}{m_{ps}^{2}}
\bar{u}({\bf q'},\lambda_{1}')  \gamma^{5}\gamma^{\mu}i(q'-q)_{\mu}
 u({\bf q},\lambda_{1})
\bar{u}({\bf -q'},\lambda_{2}')  \gamma^{5}\gamma^{\mu}i(q'-q)_{\mu}
 u({\bf -q},\lambda_{2})
\nonumber \\  &  &  /[({\bf q'-q})^{2}+m_{ps}^{2}]
\nonumber \\ & = &
\frac{f^{2}_{ps}}{(2\pi)^3} \frac{4 M^{2}}{m_{ps}^{2}}
\{\bar{u}({\bf q'},\lambda_{1}')  \gamma^{5} u({\bf q},\lambda_{1})
\bar{u}({\bf -q'},\lambda_{2}')  \gamma^{5} u({\bf -q},\lambda_{2})
\nonumber \\ & &
+[(E'-E)/(2M)]^{2}
\bar{u}({\bf q'},\lambda_{1}') \gamma^{5} \gamma^{0} u({\bf q},\lambda_{1})
\bar{u}({\bf -q'},\lambda_{2}') \gamma^{5} \gamma^{0} u({\bf -q},\lambda_{2})
\nonumber \\ & &
+[(E'-E)/(2M)]
[\bar{u}({\bf q'},\lambda_{1}')  \gamma^{5} u({\bf q},\lambda_{1})
\bar{u}({\bf -q'},\lambda_{2}')  \gamma^{5}\gamma^{0} u({\bf -q},\lambda_{2})
\nonumber \\ & &
+\bar{u}({\bf q'},\lambda_{1}')  \gamma^{5}\gamma^{0} u({\bf q},\lambda_{1})
\bar{u}({\bf -q'},\lambda_{2}')  \gamma^{5} u({\bf -q},\lambda_{2})]\}
\nonumber \\  &  &  /[({\bf q'-q})^{2}+m_{ps}^{2}] ; 
\label{31.7}\\
\nonumber \\
\lefteqn{\hspace{-.6cm}\langle {\bf q'} \lambda_{1}'\lambda_{2}'|V^{OBE}_{s}|
{\bf q}\lambda_{1}\lambda_{2}\rangle}\nonumber \\
 & = & -\frac{g^{2}_{s}}{(2\pi)^3}
\bar{u}({\bf q'},\lambda_{1}')    u({\bf q},\lambda_{1})
\bar{u}({\bf -q'},\lambda_{2}')   u({\bf -q},\lambda_{2})
  /[({\bf q'-q})^{2}+m_{s}^{2}] ;
\label{31.8}\\
\nonumber \\
\lefteqn{\hspace{-.6cm}\langle {\bf q'} \lambda_{1}'\lambda_{2}'|V^{OBE}_{v}|
{\bf q}\lambda_{1}\lambda_{2}\rangle}\nonumber \\
 &  = &
\frac{1}{(2\pi)^3} \{g_{v}
\bar{u}({\bf q'},\lambda_{1}')  \gamma_{\mu} u({\bf q},\lambda_{1})
+\frac{f_{v}}{2M}
\bar{u}({\bf q'},\lambda_{1}')
\sigma_{\mu\nu}i(q'-q)^{\nu}
 u({\bf q},\lambda_{1})\}
\nonumber \\ & & \times
\{g_{v}\bar{u}({\bf -q'},\lambda_{2}')
  \gamma^{\mu} u({\bf -q},\lambda_{2})
-\frac{f_{v}}{2M}\bar{u}({\bf -q'},\lambda_{2}')
\sigma^{\mu\nu}i(q'-q)_{\nu}
   u({\bf -q},\lambda_{2})\}
\nonumber \\ 
& &  /[({\bf q'-q})^{2}+m_{v}^{2}]
\nonumber \\ & = &
\frac{1}{(2\pi)^3} \{(g_{v}+f_{v})
\bar{u}({\bf q'},\lambda_{1}')  \gamma_{\mu} u({\bf q},\lambda_{1})
\nonumber \\ & &
-\frac{f_{v}}{2M}
\bar{u}({\bf q'},\lambda_{1}')
[(q'+q)_{\mu}+(E'-E)(g^{0}_{\mu}-\gamma_{\mu}\gamma^{0})]
 u({\bf q},\lambda_{1})\}
\nonumber \\ & & \times
\{(g_{v}+f_{v})\bar{u}({\bf -q'},\lambda_{2}')
  \gamma^{\mu} u({\bf -q},\lambda_{2})
\nonumber \\ & &
-\frac{f_{v}}{2M}\bar{u}({\bf -q'},\lambda_{2}')
[(q'+q)_{\mu}+(E'-E)(g^{\mu 0}-\gamma^{\mu}\gamma^{0})]
   u({\bf -q},\lambda_{2})\}
\nonumber \\  &  &  /[({\bf q'-q})^{2}+m_{v}^{2}] .
\label{31.9}
\end{eqnarray}
\normalsize
Working in the two-nucleon c.m. frame,
the momenta of the two incoming (outgoing) nucleons are
${\bf q}$ and $-{\bf q}$ (${\bf q'}$ and $-{\bf q'}$). 
$E\equiv\sqrt{M^{2}+{\bf q}^{2}}$ and $E'\equiv\sqrt{M^{2}+{\bf q'}^{2}}$.
Using the BbS or Thompson equation, the four-momentum transfer between
the two nucleons is $(q'-q)=(0,{\bf q'-q})$.
The Dirac equation is applied repeatedly in the
evaluations of the
 $pv$-coupling,
 the Gordon
identity~\cite{BD64} is used in the
case of the $v$-coupling.
(Note that in Eq.~(\ref{31.9}), second line from the bottom,
the term $(q'+q)_{\mu}$ carries $\mu$ as a {\it sub}script to ensure the
correct sign of the space component of that term.)
The propagator for vector bosons is
\begin{equation}
i\frac{-g_{\mu\nu}+
(q'-q)_{\mu}(q'-q)_{\nu}
/m_{v}^{2}}{-({\bf q'-q})^{2}-m_{v}^{2}}
\label{31.10}
\end{equation}
where we drop the $(q'-q)_{\mu}(q'-q)_{\nu}$-term
 which vanishes on-shell, anyhow, since the nucleon current
is conserved.
The off-shell effect of this term was examined
in Ref.~\cite{HM1} and 
found to be unimportant.

The Dirac spinors in helicity representation are given by
\begin{eqnarray}
u({\bf q},\lambda_1)&=&\sqrt{\frac{E+M}{2M}}
\left( \begin{array}{c}
       1\\
       \frac{2\lambda_1 q}{E+M}
       \end{array} \right)
|\lambda_1\rangle
\label{31.11} \\
u(-{\bf q},\lambda_2)&=&\sqrt{\frac{E+M}{2M}}
\left( \begin{array}{c}
       1\\
       \frac{2\lambda_2 q}{E+M}
       \end{array} \right)
|\lambda_2\rangle
\end{eqnarray}
with
\begin{equation}
|\lambda_1\rangle= \chi_{\lambda_1}\; , \;\;\;\;
|\lambda_2\rangle= \chi_{-\lambda_2}\; ,
\end{equation}
where $\chi$ denotes the conventional Pauli spinor.

We normalize Dirac spinors covariantly, that is
\begin{equation}
\bar{u}({\bf q},\lambda) u({\bf q},\lambda)=1,
\label{31.12}
\end{equation}
with $\bar{u}=u^{\dagger}\gamma^{0}$.

At each meson-nucleon vertex, a form factor is applied which has the 
analytical form
\begin{equation}
{\cal F}_\alpha[({\bf q}'-{\bf q})^2]=\left(\frac{\Lambda^2_\alpha-m^2_\alpha}
{\Lambda^2_\alpha+({\bf q}'-{\bf q})^2} \right)^{n_\alpha}
\label{31.13}
\end{equation}
with $m_\alpha$ the mass of the meson involved, $\Lambda_\alpha$ the so-called 
cutoff mass, and $n_\alpha$ an exponent.
Thus, the OBE amplitudes Eqs.~(\ref{31.7})-(\ref{31.9}) are multiplied
by ${\cal F}_\alpha^2$.

In practice it is desirable to have the potential represented in partial
waves, since scattering phase shifts are defined in such a representation
and nuclear structure calculations are conventionally performed
in an $LSJ$ basis. We will turn to this in the next subsection.

\subsection{Partial Wave Decomposition}

The OBE amplitudes are decomposed into partial waves according to
\begin{eqnarray}
\langle \lambda_1' \lambda_2' |V^J(q',q)| \lambda_1 \lambda_2 \rangle =
2\pi \int^{+1}_{-1} d(\cos \theta)\: d^J_{\lambda_1-\lambda_2, 
\lambda_1'-\lambda_2'}(\theta)
\langle {\bf q}'\lambda_1'\lambda_2'|V|{\bf q}\lambda_1\lambda_2\rangle
\nonumber \\ \label{32.1}
\end{eqnarray}
where $\theta$ is the angle between ${\bf q}$ and ${\bf q}'$;
$d^J_{m,m'}(\theta)$ are the conventional reduced rotation matrices.

The $d^J_{m,m'}(\theta)$ are expressed in terms of Legendre 
polynominals, $P_J(\cos \theta)$. 
The following types of integrals will occur repeatedly:
\begin{eqnarray}
\nonumber \\ 
I^{(0)}_J & \equiv &
\int_{-1}^{+1} dt\: \frac{P_J(t)}{({\bf q}'-{\bf q})^2+m^2_\alpha}
\: {\cal F}_\alpha^2[({\bf q}'-{\bf q})^2]
\nonumber \\ 
\nonumber \\ 
I^{(1)}_J & \equiv &
\int_{-1}^{+1} dt\: \frac{ t P_J(t)}{({\bf q}'-{\bf q})^2+m^2_\alpha}
\: {\cal F}_\alpha^2[({\bf q}'-{\bf q})^2]
\nonumber \\ 
\nonumber \\
I^{(2)}_J & \equiv & \frac{1}{J+1}
\int_{-1}^{+1} dt\: \frac{J t P_J(t)+P_{J-1}(t)}{({\bf q}'-{\bf q})^2+m^2_\alpha}
\: {\cal F}_\alpha^2[({\bf q}'-{\bf q})^2]
\nonumber \\ 
\nonumber \\
I^{(3)}_J & \equiv & \sqrt{\frac{J}{J+1}}
\int_{-1}^{+1} dt\: \frac{ t P_J(t)-P_{J-1}(t)}{({\bf q}'-{\bf q})^2+m^2_\alpha}
\: {\cal F}_\alpha^2[({\bf q}'-{\bf q})^2]
\label{32.2} \\
\nonumber \\ 
I^{(4)}_J & \equiv &
\int_{-1}^{+1} dt\: \frac{ t^2 P_J(t)}{({\bf q}'-{\bf q})^2+m^2_\alpha}
\: {\cal F}_\alpha^2[({\bf q}'-{\bf q})^2]
\nonumber \\ 
\nonumber \\
I^{(5)}_J & \equiv & \frac{1}{J+1}
\int_{-1}^{+1} dt\: \frac{J t^2 P_J(t) + t P_{J-1}(t)}
{({\bf q}'-{\bf q})^2+m^2_\alpha}
\: {\cal F}_\alpha^2[({\bf q}'-{\bf q})^2]
\nonumber \\ 
\nonumber \\
I^{(6)}_J & \equiv & \sqrt{\frac{J}{J+1}}
\int_{-1}^{+1} dt\: \frac{ t^2 P_J(t) - t P_{J-1}(t)}
{({\bf q}'-{\bf q})^2+m^2_\alpha}
\: {\cal F}_\alpha^2[({\bf q}'-{\bf q})^2]
\nonumber \\ 
\nonumber
\end{eqnarray}
where $t\equiv\cos \theta$. Notice that these integrals are functions of
$q',q,m_\alpha,\Lambda_\alpha, \mbox{and } n_\alpha$. 

We state the final expressions for the partial wave OBE amplitudes
in terms of the combinations of
helicity amplitudes defined in Eq.~(\ref{22.4}). More details concerning 
their derivation are to be found in Appendix E of Ref.~\cite{MHE87}.
\\ \\
Pseudo-scalar bosons ($\eta$ and $\pi$ meson; for $\pi$ apply an 
additional factor of 
{\boldmath ${\bf \tau}_1 \cdot {\bf \tau}_2$}):\\
Pseudo-vector coupling (pv)
\begin{eqnarray}
^0V^J_{pv} & = &
C_{pv}\: ( F^{(0)}_{pv}\: I^{(0)}_J + F^{(1)}_{pv}\: I^{(1)}_J )
\nonumber \\
^1V^J_{pv} & = &
C_{pv}\: ( -F^{(0)}_{pv}\: I^{(0)}_J - F^{(1)}_{pv}\: I^{(2)}_J )
\nonumber \\
^{12}V^J_{pv} & = &
C_{pv}\: ( F^{(1)}_{pv}\: I^{(0)}_J + F^{(0)}_{pv}\: I^{(1)}_J )
\label{32.3} \\
^{34}V^J_{pv} & = &
C_{pv}\: ( -F^{(1)}_{pv}\: I^{(0)}_J - F^{(0)}_{pv}\: I^{(2)}_J )
\nonumber \\
^{55}V^J_{pv} & = &
C_{pv}\:  F^{(2)}_{pv}\: I^{(3)}_J
\nonumber \\
^{66}V^J_{pv} & = &
-C_{pv}\:  F^{(2)}_{pv}\: I^{(3)}_J
\nonumber
\end{eqnarray}
with
\begin{eqnarray}
F^{(0)}_{pv} & = & E'E-M^2 + (E'-E)^2(E'E+3M^2)/(4M^2) \nonumber \\
F^{(1)}_{pv} & = & -q'q + q'q(E'-E)^2/(4M^2) \label{32.6} \\
F^{(2)}_{pv} & = & - (E'-E)[\begin{array}{c} \frac14 (E'-E)^2
+ E'E \end{array}]/M    \nonumber
\end{eqnarray}
and
\begin{equation}
C_{pv}= \frac{f^2_{ps}}{4\pi}\: \frac{2}{\pi m^2_\alpha}
\: .
\label{32.7}
\end{equation}
For the coupling constant, the following relation holds:
\begin{equation}
g_{ps}=f_{ps}\: \frac{2M}{m_{ps}}\: .
\label{32.8}
\end{equation}
\\ \\
Scalar coupling (s) ($\sigma$ and $\delta$ boson; for $\delta$ apply an 
additional factor of 
{\boldmath ${\bf \tau}_1 \cdot {\bf \tau}_2$}
):
\begin{eqnarray}
^0V^J_{s} & = &
C_{s}\: ( F^{(0)}_{s}\: I^{(0)}_J + F^{(1)}_{s}\: I^{(1)}_J )
\nonumber \\
^1V^J_{s} & = &
C_{s}\: ( F^{(0)}_{s}\: I^{(0)}_J + F^{(1)}_{s}\: I^{(2)}_J )
\nonumber \\
^{12}V^J_{s} & = &
C_{s}\: ( F^{(1)}_{s}\: I^{(0)}_J + F^{(0)}_{s}\: I^{(1)}_J )
\label{32.9} \\
^{34}V^J_{s} & = &
C_{s}\: ( F^{(1)}_{s}\: I^{(0)}_J + F^{(0)}_{s}\: I^{(2)}_J )
\nonumber \\
^{55}V^J_{s} & = &
C_{s}\:  F^{(2)}_{s}\: I^{(3)}_J
\nonumber \\
^{66}V^J_{s} & = &
C_{s}\:  F^{(2)}_{s}\: I^{(3)}_J
\nonumber
\end{eqnarray}
with
\begin{equation}
C_{s}= \frac{g^2_{s}}{4\pi}\: \frac{1}{2\pi M^2}
\label{32.10}
\end{equation}
and
\begin{eqnarray}
F^{(0)}_{s} & = & - (E'E+M^2) \nonumber \\
F^{(1)}_{s} & = & q'q \label{32.11} \\
F^{(2)}_{s} & = & M(E'+E) .  \nonumber
\end{eqnarray}

\begin{table}[t]
\centering
\scriptsize
\caption{Parameters of the relativistic OBEPs used in this work.}
\begin{tabular}{llllllll}
\\ \hline\hline \\
  &   & \multicolumn{2}{l}{\it Potential A}
& \multicolumn{2}{l}{\it Potential B } &
\multicolumn{2}{l}{\it Potential C}  \\
\multicolumn{2}{l}{\it Meson Parameters}\\
  & $m_{\alpha}$(MeV)
     &  $g^{2}_{\alpha}/4\pi$
     &  $\Lambda_{\alpha}$(GeV)
     &  $g^{2}_{\alpha}/4\pi$
     &  $\Lambda_{\alpha}$(GeV)
     &  $g^{2}_{\alpha}/4\pi$
     &  $\Lambda_{\alpha}$(GeV) \\
\\ \hline \\

$\pi$ & 138.03  & 14.9 & 1.05 & 14.6 & 1.2 & 14.6 & 1.3  \\

$\eta$ & 548.8  &  7   & 1.5 &   5  & 1.5 & 3 & 1.5 \\

$\rho$ & 769   & 0.99 & 1.3 &  0.95 & 1.3 & 0.95 & 1.3 \\

$\omega$ & 782.6 & 20 & 1.5 & 20    &  1.5 & 20 & 1.5  \\

$\delta$ & 983  & 0.7709 & 2.0 & 3.1155 &   1.5 & 5.0742 & 1.5 \\

$\sigma$ & 550  & 8.3141 & 2.0 &  8.0769  &  2.0
 & 8.0279 & 1.8 \\

\\ \hline \\
{\it Deuteron}\\
 $-\epsilon_{d}$ (MeV)&&  2.22459 && 2.22468 && 2.22450 & [2.224575(9)]\\
 $P_{D}$ (\%) && 4.47 && 5.10 && 5.53 & [---]\\
 $Q_{d}$ (fm$^{2}$) && 0.274 $^{a}$ && 0.279 $^{a}$ && 0.283 $^{a}$ & 
[0.2860(15)]\\
 $\mu_{d}$ ($\mu_{N}$) && 0.8543 $^{a}$ && 0.8507 $^{a}$ && 0.8482 $^{a}$ & 
0.857406(1)]\\
 $A_{S}$ (fm$^{-1/2}$) && 0.8984 && 0.8968 && 0.8971 & [0.8846(8)]\\
 $D/S$ &&   0.0255 &&   0.0257  && 0.0260 & [0.0264(12)]\\ \\
\multicolumn{2}{l}{\it Low Energy Scattering}\\
 $a_{np}$ (fm) && -23.752 &&  -23.747 && -23.740 & [-23.748(10)]\\
 $r_{np}$ (fm) && 2.69    &&  2.67 && 2.66 & [2.75(5)]\\
 $a_{t}$ (fm) &&      5.482   &&  5.474 && 5.475 & [5.419(7)]\\
 $r_{t}$ (fm) && 1.829 && 1.819 && 1.821 & [1.754(8)]\\
\\ \hline\hline \\
\multicolumn{8}{l}{\footnotesize
The meson parameters define the potentials. The deuteron and low energy
scattering}
\\ \multicolumn{8}{l}{\footnotesize
parameters are predictions
 by these potentials. The experimental values are given in square}
\\ \multicolumn{8}{l}{\footnotesize
brakets. For notation and references to the empirical values see 
Table 4.2 of Ref.~[10]..}
\\ \multicolumn{8}{l}{\footnotesize
It is always used:
 $f_{\rho}/g_{\rho} = 6.1$ and $f_{\omega}/g_{\omega}=0.0$.}
\\ \multicolumn{8}{l}{\footnotesize
$^{a}$ Meson-exchange current contributions not included.}
\end{tabular}
\end{table}

\begin{figure}[t]
\centerline{\psfig{figure=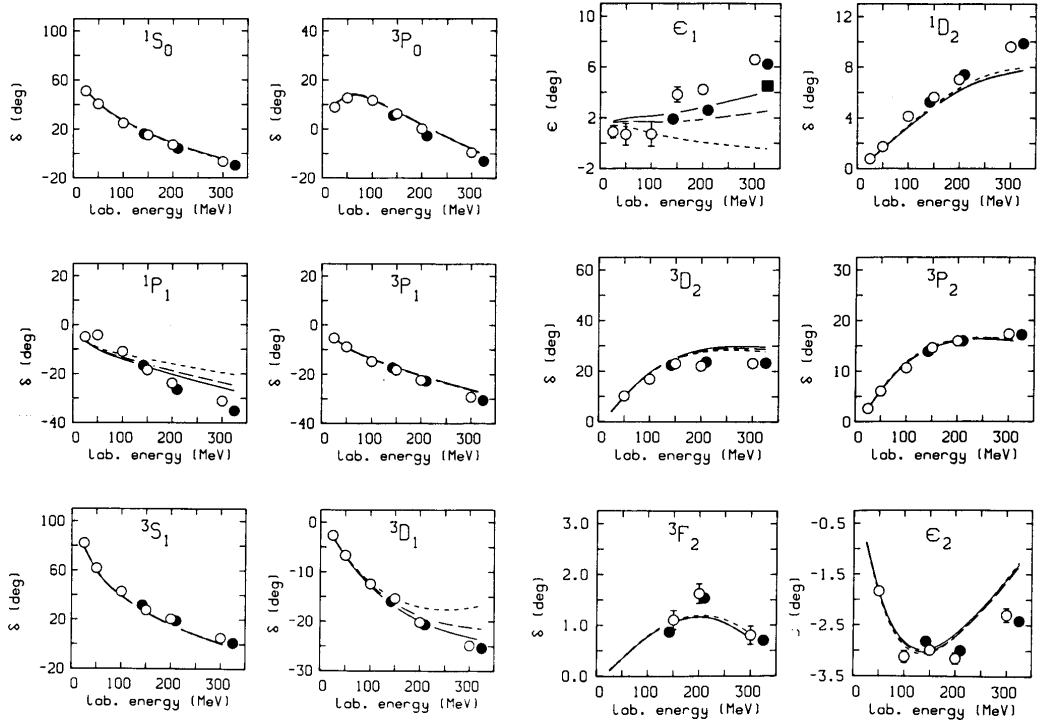,width=14cm}}              
\caption{Neutron-proton phase shifts as predicted by the three
potentials defined in Table 2. The solid, dashed, and dotted lines
refer to Potentials C, B, and A, respectively.
The open circles represent the phase shift analysis by Arndt
{\it et al.} [70]; the full dots and the full square
are from two analyses by Bugg, Refs.~[71] and [72],
respectively.}
\end{figure}

Vector bosons (v) ($\omega$ and $\rho$ meson; for $\rho$ apply an 
additional factor of 
{\boldmath ${\bf \tau}_1 \cdot {\bf \tau}_2$}
):\\
Vector-vector coupling
\begin{eqnarray}
^0V^J_{vv} & = &
C_{vv}\: (2E'E-M^2)\: I^{(0)}_J 
\nonumber \\
^1V^J_{vv} & = &
C_{vv}\: ( E'E\: I^{(0)}_J + q'q\: I^{(2)}_J )
\nonumber \\
^{12}V^J_{vv} & = &
C_{vv}\: ( 2q'q\: I^{(0)}_J + M^2\: I^{(1)}_J )
\label{32.12} \\
^{34}V^J_{vv} & = &
C_{vv}\: ( q'q\: I^{(0)}_J + E'E\: I^{(2)}_J )
\nonumber \\
^{55}V^J_{vv} & = &
-C_{vv}\:  ME\: I^{(3)}_J
\nonumber \\
^{66}V^J_{vv} & = &
-C_{vv}\:  ME'\: I^{(3)}_J
\nonumber
\end{eqnarray}
with
\begin{equation}
C_{vv}= \frac{g^2_{v}}{4\pi}\: \frac{1}{\pi M^2}\: .
\label{32.13}
\end{equation}
Tensor-tensor coupling
\begin{eqnarray}
^0V^J_{tt} & = &
C_{tt}\: \{ (q'^2+q^2)(3E'E+M^2)\: I^{(0)}_J \nonumber \\
 & & + [q'^2+q^2-2(3E'E+M^2)]q'q\: I^{(1)}_J - 2q'^2q^2\: I^{(4)}_J \}
\nonumber \\
^1V^J_{tt} & = &
C_{tt}\: \{ [4q'^2q^2+(q'^2+q^2)(E'E-M^2)]\: I^{(0)}_J \nonumber \\
 & & + 2(E'E+M^2)q'q\: I^{(1)}_J \nonumber \\
 & & - (q'^2+q^2+4E'E)q'q\: I^{(2)}_J - 2q'^2q^2\: I^{(5)}_J \}
\label{32.14} \\
^{12}V^J_{tt} & = &
C_{tt}\: \{ [4M^2-3(q'^2+q^2)]q'q\: I^{(0)}_J \nonumber \\
 & & + [6q'^2q^2-(q'^2+q^2)(E'E+3M^2)]\: I^{(1)}_J +
  2(E'E+M^2)q'q\: I^{(4)}_J \}
\nonumber \\
^{34}V^J_{tt} & = &
C_{tt}\: \{ -(q'^2+q^2+4E'E)q'q\: I^{(0)}_J - 2q'^2q^2\: I^{(1)}_J
\nonumber \\
 & & +[4q'^2q^2+(q'^2+q^2)(E'E-M^2)]\: I^{(2)}_J
     +2(E'E+M^2)q'q\: I^{(5)}_J \}
\nonumber \\
^{55}V^J_{tt} & = &
C_{tt}\:  M \{ [ E'(q'^2+q^2) + E (3q'^2 - q^2)]\: I^{(3)}_J
 - 2(E'+E)q'q\: I^{(6)}_J \}
\nonumber \\
^{66}V^J_{tt} & = &
C_{tt}\:  M \{ [ E(q'^2+q^2) + E'(3q^2 - q'^2)]\: I^{(3)}_J
 - 2(E'+E)q'q\: I^{(6)}_J \}
\nonumber
\end{eqnarray}
with
\begin{equation}
C_{tt}= \frac{f^2_{v}}{4\pi M^2}\: \frac{1}{8\pi M^2}
\label{32.15}
\end{equation}
Vector-tensor coupling
\begin{eqnarray}
^0V^J_{vt} & = &
C_{vt}\: M [ (q'^2+q^2)\: I^{(0)}_J - 2q'q\: I^{(1)}_J ]
\nonumber \\
^1V^J_{vt} & = &
C_{vt}\: M [ -(q'^2+q^2)\: I^{(0)}_J + 2q'q\: I^{(2)}_J ]
\nonumber \\
^{12}V^J_{vt} & = &
C_{vt}\: M [ 6q'q\: I^{(0)}_J - 3(q'^2+q^2)\: I^{(1)}_J ]
\label{32.16} \\
^{34}V^J_{vt} & = &
C_{vt}\: M [ 2q'q\: I^{(0)}_J - (q'^2+q^2)\: I^{(2)}_J ]
\nonumber \\
^{55}V^J_{vt} & = &
C_{vt}\:  (E'q^2+3Eq'^2)\: I^{(3)}_J
\nonumber \\
^{66}V^J_{vt} & = &
C_{vt}\:  (Eq'^2+3E'q^2)\: I^{(3)}_J
\nonumber
\end{eqnarray}
with
\begin{equation}
C_{vt}= \frac{g_{v}f_{v}}{4\pi M}\: \frac{1}{2\pi M^2}
\label{32.17}
\end{equation}

We use units such that $\hbar=c=1$.
Energies, masses and momenta are in  MeV.
The potential is in units of MeV$^{-2}$.
 The conversion factor is $\hbar c 
=197.327$ MeV fm.  If the user wants to relate our units 
and conventions to those used in the work of other researchers, he/she should 
compare our Eq.~(\ref{22.5}) and our phase shift relation Eq.~(\ref{23.1}) 
with the corresponding equations in other works.
A computer code for the relativistic OBE potential is published
in Ref.~\cite{Mac93}.

We note that when, in the nuclear medium (nuclear matter),
the free Dirac spinors, Eq.~(69),
are replaced by the in-medium Dirac spinors, Eq.~(92),
then all $E$ and $E'$ in the above expressions
are replaced by $\tilde{E}$ and $\tilde{E}'$, respectively,
and all $M$ by $\tilde{M}$; an exception from this rule are the factors
$C_{tt}$, Eq.~(86), and $C_{vt}$, Eq.~(88), which are in the medium:
\begin{displaymath}
C_{tt}= \frac{f^2_{v}}{4\pi M^2}\: \frac{1}{8\pi \tilde{M}^2} \; ,
\: \: \: \: \; \; \; \; \; \; \; \; \; 
C_{vt}= \frac{g_{v}f_{v}}{4\pi M}\: \frac{1}{2\pi \tilde{M}^2} \; ;
\end{displaymath}
notice that the factor $f_v/M$ comes from the Lagrangian, Eq.~(63),
and we assume here that it does not change in the medium.

\subsection{Meson Parameters and Two-Nucleon Predictions}

We give here three
examples for relativistic momentum-space OBEPs, 
which have become known as the Bonn A, B,
and C potentials\footnote{It is customary to 
denote the OBE parametrizations of the Bonn full model~\cite{MHE87} by the 
first letters of the alphabet, A, B or C, see e.~g. Appendix A of 
Ref.~\cite{Mac89}.}.
They are presented in Appendix A, Table A.2, of Ref.~\cite{Mac89}.
These potentials use the pseudovector coupling for $\pi$ 
and $\eta$ and are constructed in the framework of the Thompson
equation, Eqs.~(27-30). 

In Table~2 we 
give the meson parameters as well as
the predictions for the deuteron and the 
low-energy scattering parameters.
Phase shifts for 
neutron-proton scattering are displayed in Fig.~2.

\section{The Dirac-Brueckner Approach}

As mentioned in the Introduction, the essential point of the
Dirac-Brueckner approach is  to use  the Dirac
equation for the single-particle motion in nuclear matter
\begin{equation}
(\not\!p-M-U)\tilde{u}({\bf p},s)=0
\end{equation}
or in Hamiltonian form
\begin{equation}
(\mbox{\boldmath $\alpha$} \cdot {\bf p} + \beta M + \beta U)
\tilde{u}({\bf p},s) = \epsilon_{p} \tilde{u}({\bf p},s)
\end{equation}
with
\begin{equation}
U=U_{S} +\gamma^{0}U_{V}
\end{equation}
where $U_{S}$ is an attractive scalar
and $U_{V}$ (the time-like component of) a repulsive vector field.
(Notation as in Ref. \cite{BD64}; $\beta=\gamma^{0}$,
  $\alpha^{l}=\gamma^{0}\gamma^{l}$.)
$M$ is the mass of the free nucleon.

The fields, $U_{S}$ and $U_{V}$,
 are in the order of several hundred MeV and strongly density dependent
(numbers will be given below).
In nuclear matter they can be determined self-consistently.
The resulting fields are in close agreement
with those obtained in the Dirac phenomenology of scattering.

The solution of Eq.\ (89) is
\begin{equation} \tilde{u}({\bf p},s)=\sqrt{\frac{\tilde{E}_{p}+\tilde{M}}{2\tilde{M}}} \left( \begin{array}{c} 1\\ \frac{\mbox{\boldmath $\sigma \cdot  p$}}{\tilde{E}_{\bf p}+\tilde{M}} \end{array} \right) \chi_{s}
\end{equation}
with
\begin{equation}
\tilde{M}=M+U_{S},
\end{equation}
\begin{equation}
\tilde{E}_{\bf p}=\sqrt{\tilde{M}^{2}+{\bf p}^{2}},
\end{equation}
and $\chi_{s}$ a Pauli spinor.
The covariant normalization is
 $\bar{\tilde{u}}({\bf p},s)\tilde{u}({\bf p},s)=1$.
Notice that the Dirac spinor Eq.~(92) is obtained
 from the free Dirac spinor by replacing $M$ by $\tilde{M}$. 

As in conventional Brueckner theory, the basic quantity
in the Dirac-Brueckner approach is a
$G$-matrix, which satisfies an integral equation.
In this relativistic approach, a relativistic
three-dimensional equation is chosen.
Following the basic philosophy of traditional Brueckner theory,
this equation is applied to nuclear matter in strict
analogy to free scattering.

Including the necessary medium effects, the Thompson equation in the nuclear 
matter rest frame reads [cf.\ Eq.~(26)]
\begin{equation}
\tilde{G}({\bf q', q|P},\tilde{z})=\tilde{V}({\bf q'},{\bf q})+
\int \frac{d^{3}k}{(2\pi)^{3}}
 \tilde{V}({\bf q'}, {\bf k}) \frac{\tilde{M}^{2}}
{\tilde{E}_{\frac{1}{2}{\bf P}+{\bf k}}^{2}}
 \frac{Q({\bf k,P})}{\tilde{z}-2\tilde{E}_{\frac{1}{2}{\bf P}+{\bf k}}}
\tilde{G}({\bf k,q|P},\tilde{z})
\end{equation}
with
\begin{equation}
\tilde{z}=2\tilde{E}_{\frac{1}{2}{\bf P}+{\bf q}}.
\end{equation}
${\bf P}$ is the c.m. momentum, and ${\bf q, k}$ and ${\bf q'}$ are
the initial, intermediate and final relative momenta, respectively,
of the two particles interacting in nuclear matter.
The Pauli operator $Q$ projects onto unoccupied states.
 In Eq. (95) we suppressed
 the $k_{F}$ dependence as well as spin (helicity) and isospin
indices.
For $|\frac{1}{2}{\bf P\pm q}|$ and $|\frac{1}{2}{\bf P\pm k}|$
the angle average
is used.

\begin{figure}[t]
\centerline{\psfig{figure=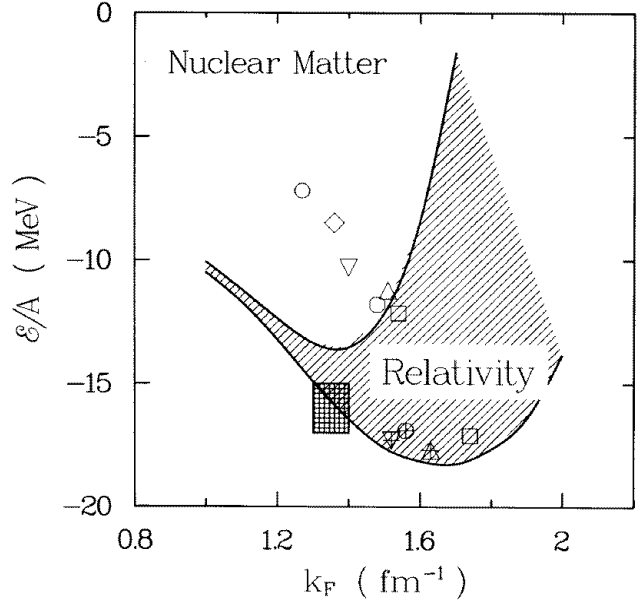,width=8.5cm}}
\caption{The repulsive relativistic effect in nuclear matter
as obtained in a Dirac-Brueckner-Hartree-Fock calculation
using Potential B. Saturation points from conventional calculations
are displayed in the background (cf.\ Fig.~1). The checked rectangle
represents the empirical value for nuclear matter saturation.}
\end{figure}

\begin{table}[t]
\footnotesize
\centering
\caption{Results of a relativistic Dirac-Brueckner calculation
in comparison to the corresponding non-relativistic one using Potential B.}
\begin{tabular}{crrrrrrrrrr}
\\  \hline\hline \\
 &  \multicolumn{5}{c}{\footnotesize\it relativistic} & &
 \multicolumn{4}{c}{\footnotesize\it
 non-relativistic} \\
\\  \cline{2-6} \cline{8-11}
\\ \multicolumn{1}{c}{$k_{F}$}  &
\multicolumn{1}{c}{ ${\cal E}/A$ } &
 $\tilde{M}/M$ &
\multicolumn{1}{c}{ $U_{S}$} &
\multicolumn{1}{c}{ $U_{V}$} &
\multicolumn{1}{c}{ $\kappa$} & \makebox[2ex]{} &
\multicolumn{1}{c}{ ${\cal E}/A$}  &
$\tilde{M}/M$ &
\multicolumn{1}{c}{ $U_{0}$$^{a}$} &
\multicolumn{1}{c}{ $\kappa$} \\
\multicolumn{1}{c}{(fm$^{-1}$)} &
(MeV)  &  & (MeV) & (MeV) & (\%) & & (MeV) & & (MeV) & (\%) \\ \\
 \hline \\
0.8 &  --7.02 & 0.855 & --136.2 & 104.0 & 23.1 & &--7.40&0.876&--33.0&26.5\\
0.9 &  --8.58 & 0.814 & --174.2 & 134.1 & 18.8 & &--9.02&0.836&--41.0&21.6\\
1.0 & --10.06 & 0.774 & --212.2 & 164.2 & 16.1 & &--10.49&0.797&--49.0&18.5\\
1.1 & --11.18 & 0.732 & --251.3 & 195.5 & 12.7 & & --11.69 & 0.760 & --58.1
 & 14.2 \\
1.2 & --12.35 & 0.691 & --290.4 & 225.8 & 11.9 & & --13.21 & 0.725 & --68.5
 & 12.9 \\
1.3 & --13.35 & 0.646 & --332.7 & 259.3 & 12.5 & & --14.91 & 0.687 & --80.5
 & 13.1\\
1.35 & --13.55 & 0.621 & --355.9 & 278.4 & 13.0 & & --15.58 & 0.664 & --86.8
 & 13.2\\
1.4 & --13.53 & 0.601 & --374.3 & 293.4 & 13.8 & & --16.43 & 0.651 & --93.2
 & 13.5\\
1.5 & --12.15 & 0.559 & --413.6 & 328.4 & 14.4 & & --17.61 & 0.618 & --106.1
 & 13.0\\
1.6 &  --8.46 & 0.515 & --455.2 & 371.0 & 15.8 & & --18.14 & 0.579 & --119.4
 & 12.7\\
1.7 &  --1.61 & 0.477 & --491.5 & 415.1 & 18.4 & & --18.25 & 0.545 & --133.2
 & 13.2\\
1.8 &  +9.42 & 0.443 & --523.4 & 463.6 & 21.9 & & --17.65 & 0.489 & --147.2
 & 14.3\\
1.9 &  25.26 & 0.418 & --546.7 & 513.5 & 25.2 & & --16.41 & 0.480 & --160.7 &
 15.0\\
2.0 &  47.56 & 0.400 & --563.6 & 568.6 & 27.5 & & --13.82 & 0.449 & --173.6
 & 15.3\\
2.1 &  77.40 & 0.381 & --581.3 & 640.9 & 30.2 & &  --9.70 & 0.411 & --186.3
 & 15.7\\
2.2 & 114.28 & 0.370 & --591.2 & 723.5 & 33.3 & &  --3.82 & 0.373 & --198.1
 & 16.3\\
\\ \hline\hline  \\
\multicolumn{11}{l}{\footnotesize
As a funtion of the  Fermi momentum
$k_{F}$, we list:
 the energy per nucleon ${\cal E}/A$,}
\\ \multicolumn{11}{l}{\footnotesize
 $\tilde{M}/M$, the single-particle
scalar and vector potentials $U_{S}$ and $U_{V}$, and the wound}
\\ \multicolumn{11}{l}{\footnotesize
  integral
$\kappa$.}
\\ \multicolumn{11}{l}{\footnotesize
$^{a}$ $U_{0}$ is to be compared to $U_{S}+U_{V}$, cf.\ Eq. (104).}
\end{tabular}
\end{table}

\begin{table}[t]
\footnotesize
\centering
\caption{Partial wave contributions to the energy
 in nuclear matter (in MeV) for a non-relativistic
and a relativistic calculation using Potential B.}
\begin{tabular}{crrrrrrrr}
\\ \hline\hline  \\
  & \multicolumn{2}{c}{$k_{F}=1.1$ fm$^{-1}$} & &
 \multicolumn{2}{c}{$k_{F}=1.35$ fm$^{-1}$} & &
 \multicolumn{2}{c}{$k_{F}=1.6$ fm$^{-1}$}  \\
\\ \cline{2-3} \cline{5-6} \cline{8-9}
\\ State & non-rel. & relativ. & \makebox[2ex]{}
 & non-rel. & relativ. & \makebox[2ex]{}
 & non-rel. & relativ. \\
\\ \hline \\
$^{1}\!S_{0}$ &
      --10.79  &
      --11.18  & &
      --16.01  &
      --16.42 & &
      --21.51 &
      --20.36   \\
$^{3}\!P_{0}$             &
       --2.07 &
       --1.48 & &
       --3.74 &
       --1.34 & &
       --5.61 &
       +2.17 \\
$^{1}\!P_{1}$             &
      1.73 &
     1.77  & &
     3.25 &
     3.45 & &
     5.33 &
     6.08 \\
$^{3}\!P_{1}$             &
       4.71  &
      5.27  & &
      9.77  &
     12.33  & &
     17.69  &
     26.65  \\
$^{3}\!S_{1}$             &
      --15.41  &
      --14.16  & &
      --20.10  &
      --17.10  & &
      --23.77  &
      --17.03  \\
$^{3}\!D_{1}$             &
      0.59 &
      0.57  & &
      1.38 &
      1.29 & &
      2.64 &
      2.25 \\
$^{1}\!D_{2}$             &
       --0.95  &
       --0.91  & &
       --2.28  &
       --2.01  & &
       --4.57  &
       --3.39 \\
$^{3}\!D_{2}$             &
       --1.70 &
       --1.62  & &
       --4.00  &
       --3.56 & &
       --7.71 &
       --5.99 \\
$^{3}\!P_{2}$             &
       --3.10 &
       --2.92  & &
       --7.06 &
       --6.28 & &
       --13.31 &
       --10.73 \\
$^{3}\!F_{2}$             &
       --0.19 &
       --0.18 & &
       --0.54 &
       --0.44 & &
       --1.19 &
       --0.67 \\
$^{1}\!F_{3}$             &
      0.32  &
      0.31 & &
      0.80 &
      0.75 & &
      1.60 &
      1.40 \\
$^{3}\!F_{3}$             &
      0.56  &
      0.55  & &
      1.51  &
      1.43  & &
      3.20  &
      2.87  \\
$^{3}\!D_{3}$             &
      --0.01  &
       0.00  & &
       --0.03  &
       0.00  & &
      --0.11  &
       --0.02  \\
$^{3}\!G_{3}$             &
       0.06  &
       0.06  & &
      0.20 &
      0.18  & &
      0.49  &
      0.41 \\
$J\geq 4$             &
       --0.34  &
       --0.33  & &
       --1.07  &
       --0.98  & &
       --2.57  &
       --2.13  \\ \\
Total\\
 potential energy             &
       --26.61  &
       --24.25 & &
       --37.93 &
       --28.72 & &
        --49.38 &
        --18.51 \\
Kinetic energy            &
       14.91  &
       13.07 & &
       22.35 &
       15.16  & &
       31.23  &
       10.05 \\
Total energy             &
       --11.69  &
       --11.18 & &
       --15.58 &
       --13.55 & &
       --18.14 &
        --8.46 \\
\\ \hline\hline \\
\end{tabular}
\end{table}

\begin{table}[t]
\footnotesize
\centering
\caption{Energy per particle, ${\cal E}/A$,
 Fermi momentum, $k_{F}$, and kompression modulus, $K$,
at saturation for nuclear matter with and without
relativistic effects.}
\begin{tabular}{ccccccccc}
\\  \hline\hline \\
 &  & \multicolumn{3}{c}{\normalsize\it relativistic} & &
 \multicolumn{3}{c}{\normalsize\it non-relativistic} \\
\\  \cline{3-5} \cline{7-9}
\\ Potential &  \makebox[2ex]{} &
\multicolumn{1}{c}{ ${\cal E}/A$ } &
\multicolumn{1}{c}{$k_{F}$}  &
\multicolumn{1}{c}{ K } & \makebox[2ex]{} &
\multicolumn{1}{c}{ ${\cal E}/A$}  &
\multicolumn{1}{c}{$k_{F}$}  &
\multicolumn{1}{c}{ K } \\
  & & (MeV) & (fm$^{-1}$) & (MeV)  &  & (MeV) & (fm$^{-1}$) &  (MeV) \\ \\
 \hline \\
A & & --15.59 & 1.40 & 290 & & --23.55 & 1.85 & 204 \\ \\
B & & --13.60 & 1.37  &  249 & & --18.30 & 1.66  &  160   \\ \\
C & & --12.26 & 1.32 & 185 & & --15.75 & 1.54 & 143
\\ \\ \hline\hline \\
\end{tabular}
\end{table}

The energy per nucleon in nuclear matter, which is the objective of these 
calculations, is considered in the nuclear matter rest frame.
 Thus, the $G$-matrix is needed for the nuclear matter rest frame.
Equation~(95) gives this nuclear matter $G$-matrix 
 directly in that rest frame.
Alternatively, one can calculate the $G$-matrix first in the two-nucleon 
center-of-mass (c.m.) system (as customary in calculations of the $\cal 
T$-matrix in two-nucleon scattering), and then perform a
 Lorentz transformation to the rest frame.
This method, which is described in detail in Ref.~\cite{HS87}, is, however, 
complicated, involved, and cumbersome. The advantage of 
our procedure is that it avoids this complication. 
Further treatments of Eq.\ (95) can follow the lines
established from conventional Brueckner theory, as e.\ g.\  the use of the
angle averaged Pauli projector etc.. Numerically the equation can be solved
by standard methodes of momentum space Brueckner calculations \cite{HT70}.

The essential difference to standard Brueckner theory is the use
of the potential
 $\tilde{V}$ in Eq.\ (95). Indicated by the tilde,
this meson-theoretic potential is evaluated
by using the spinors of Eq.\ (92) instead of the free spinors
 applied in scattering as well as
in conventional ('non-relativistic')
 Brueckner theory. Since $U _{S}$
 (and $\tilde{M}$) are strongly density dependent, so is the potential
 $\tilde{V}$.
$\tilde{M}$ decreases with density.
 The essential effect
in nuclear matter is a suppression of the
(attractive) $\sigma$-exchange; this suppression
 increases with density, providing
additional saturation. It turns out (see figures below) that this
effect is so strongly density-dependent that
 {\it the
empirical saturation and incompressibility}
can be reproduced.
Furthermore, the prediction for the Landau parameter $f_{0}$ is considerably
improved without deteriorating the other parameters (see Table 6 below).
Note, that sum rules require $f_{0} > -1$ at nuclear matter
density \cite{Mig67}.

The single-particle potential
\begin{equation}
U(m)=\frac{\tilde{M}}{\tilde{E_{m}}}
\langle m| U | m \rangle =\frac{\tilde{M}}{\tilde{E_{m}}}
 \langle m|U_{S}+\gamma^{0}U_{V}|m\rangle
=\frac{\tilde{M}}{\tilde{E_{m}}}
 U_{S} +U_{V}
\end{equation}
is the self-energy of the nucleon which
is defined in terms of the $G$-matrix formally in the usual way
\begin{equation}
U(m)=Re\sum_{n\leq k_{F}}\frac{\tilde{M}^{2}}{\tilde{E_{n}}\tilde{E_{m}}}
 \langle mn|\tilde{G}(\tilde{z})|mn-nm\rangle
\end{equation}
where $m$ denotes a state below or above the Fermi surface (continuous
choice).
The constants $U_{S}$ and $U_{V}$ are determined from Eqs.~(97) and (98).
Note that Eq.\ (91) is an approximation, since the scalar
and vector fields are in general
 momentum dependent; however, it has been shown that this momentum
dependence is very weak \cite{MB85}.

The energy per nucleon in nuclear matter is 
\begin{equation}
\frac{{\cal E}}{A}=\frac{1}{A}
\sum_{m\leq k_{F}}\frac{\tilde{M}}{\tilde{E_{m}}}
\langle m | \mbox{\boldmath $\gamma$} \cdot {\bf p}_{m}
+M|m\rangle + \frac{1}{2A} \sum_{m,n\leq k_{F}}
\frac{\tilde{M}^{2}}{\tilde{E}_{m} \tilde{E}_{n}}
\langle mn|\tilde{G}(\tilde{z})|mn-nm\rangle -M
\; .
\end{equation}
In Eqs.\ (98) and (99) we use
\begin{equation}
\tilde{z}=\tilde{E}_{m}+\tilde{E}_{n}.
\end{equation}
The expression for the energy, Eq.~(99), is denoted by
 the Dirac-Brueckner-Hartree-Fock (DBHF) approximation. If $\tilde{M}$ is 
replaced by $M$, we will speak of the Brueckner-Hartree-Fock (BHF) 
approximation, since this case, qualitatively and quantitatively, corresponds 
to conventional non-relativistic Brueckner theory. Thus, we will occasionally 
denote the DBHF calculation by `relativistic' and the BHF calculation by
 `non-relativistic' (though, strictly speaking, all our calculations are 
relativistic).  

In Equations.\ (97-99) the states $|m\rangle$ and $|n\rangle$ are
represented by Dirac spinors of the kind Eq.\ (92)
and an appropriate isospin wavefunction,
$\langle m|$ and $\langle n|$ are the adjoint Dirac spinors
$\bar{\tilde{u}}=\tilde{u}^{\dagger}\gamma^{0}$
with $\bar{\tilde{u}}\tilde{u}=1$;
$\tilde{E_{m}}\equiv\sqrt{\tilde{M}^{2}+{\bf p}^{2}_{m}}$.
The states of the nucleons in nuclear matter, $w$, are to be normalized by 
$w^{\dagger}w=1$. This is achieved by defining $w\equiv 
\sqrt{\tilde{M}/\tilde{E}}\times\tilde{u}$ which explains factors of 
$\tilde{M}/\tilde{E}$ in Eqs.~(97-99).

The first term on the r.h.s. of Eq.\ (99) --- the `kinetic
energy' --- is in more explicit form
\begin{equation}
\frac{1}{A} \sum_{m\leq k_{F}} \frac{M\tilde{M}+{\bf p}_{m}^{2}}
{\tilde{E_{m}}}.
\end{equation}
The single particle energy is
\begin{eqnarray}
\epsilon_{m}&=&\frac{\tilde{M}}{\tilde{E_{m}}}
\langle m| \mbox{\boldmath $\gamma$} \cdot {\bf p}_{m} + M |m\rangle
+ U(m)\\
 & = & \tilde{E_{m}} + U_{V} \\
 & = & \tilde{E_{m}} -\tilde{M}+M+U_{0}.
\end{eqnarray}

\section{Results for Nuclear Matter}

\begin{table}[t]
\footnotesize
\centering
\caption{Landau parameters at various densities
 from a non-relativistic and a relativistic
nuclear matter calculation using Potential B.}
\begin{tabular}{ccrrrrr}
\\ \hline\hline  \\
$k_F$ & & & & & &  \\
 (fm$^{-1}$)  & density &
 & $f_{0}$   &  $f_{0}'$   & $g_{0}$  &  $g_{0}'$  \\
\\ \hline \\
1.0 & $0.4\rho_{0}$
 & relativistic &
 --1.37 & 0.57  &  0.22  &  0.66  \\
 & & non-relativisic &
 --1.50 & 0.62  & 0.16   &0.66   \\ \\

1.35 & $\rho_{0}$
 & relativistic &
 --0.79 & 0.35  &  0.28  &  0.67 \\
 & & non-relativisic &
 --1.27 & 0.38  &  0.15  &  0.67   \\ \\
1.7 & $2\rho_{0}$
 & relativistic &
 0.56 & 0.29  &  0.36  &  0.68  \\
 & & non-relativisic &
 --0.99 & 0.20  &  0.14  &  0.69  \\ \\
2.0 & $3.25\rho_{0}$
 & relativistic &
 2.21 & 0.37  &  0.38  &  0.69  \\
 & & non-relativisic &
 --0.71 & 0.09  &  0.11  &  0.71   \\ \\
\multicolumn{3}{c}{(empirical)} & ($0\pm 0.2$) & ($\approx 0.8$) & ($\approx 
0.2$) & ($\approx 0.9$) \\ \\ 
 \hline\hline \\
\end{tabular}
\end{table}

\begin{table}[t]
\footnotesize
\centering
\caption{Nuclear matter results of
a relativistic Dirac-Brueckner calculation
applying Potential A.
Notation as in Table 3.}
\begin{tabular}{crrrrrrr}
\\  \hline\hline \\
 & \multicolumn{5}{c}{\it relativistic} &
 & \multicolumn{1}{c}{\it non-relativistic} \\
\\ \cline{2-6} \cline{8-8} \\
\multicolumn{1}{c}{$k_{F}$}  &
\multicolumn{1}{l}{ ${\cal E}/A$ } &
 $\tilde{M}/M$ &
\multicolumn{1}{c}{ $U_{S}$} &
\multicolumn{1}{c}{ $U_{V}$} &
\multicolumn{1}{c}{ $\kappa$} &&
 ${\cal E}/A$  \\
\multicolumn{1}{c}{(fm$^{-1}$)} &
 (MeV) &   & (MeV) & (MeV) & (\%)& & (MeV) \\ \\
 \hline \\
0.8 &  --7.27 & 0.857 & --134.3 & 101.6 & 22.6 & & --7.60\\
0.9 &  --8.97 & 0.817 & --172.1 & 131.1 & 18.1 & & --9.38\\
1.0 & --10.62 & 0.777 & --209.8 & 160.6 & 15.2 & & --11.01\\
1.1 & --11.96 & 0.736 & --248.5 & 191.1 & 11.5 && --12.44 \\
1.2 & --13.44 & 0.692 & --288.8 & 222.0 & 10.6 && --14.24 \\
1.3 & --14.86 & 0.647 & --331.6 & 255.0 & 11.0 && --16.35 \\
1.35 & --15.32 & 0.621 & --355.7 & 274.7 & 11.5 && --17.28 \\
1.4 & --15.59 & 0.601 & --374.9 & 289.8 & 12.1 && --18.41 \\
1.5 & --14.88 & 0.557 & --416.3 & 325.7 & 12.7 && --20.25 \\
1.6 & --11.96 & 0.511 & --459.6 & 368.6 & 14.2 && --21.64 \\
1.7 &  --5.88 & 0.470 & --497.2 & 412.8 & 16.9 && --22.76 \\
1.8 &  +4.44 & 0.435 & --530.4 & 461.6 & 20.4 && --23.38 \\
1.9 &  19.72 & 0.409 & --554.8 & 512.0 & 23.7 && --23.54 \\
2.0 &  41.62 & 0.390 & --572.4 & 567.5 & 26.3 && --22.60 \\
2.1 &  71.20 & 0.371 & --590.2 & 640.3 & 29.1 && --20.42 \\
2.2 & 106.52 & 0.366 & --595.0 & 732.5 & 31.9 && --16.74 \\
\\ \hline  \\
\end{tabular}
\end{table}

\begin{table}[t]
\footnotesize
\centering
\caption{Nuclear matter results of
a relativistic Dirac-Brueckner calculation
applying Potential C.
Notation as in Table 3.}
\begin{tabular}{crrrrrrr}
\\  \hline\hline \\
 & \multicolumn{5}{c}{\it relativistic}&
 &  \multicolumn{1}{c}{\it non-relativistic}\\
\\ \cline{2-6} \cline{8-8}  \\
\multicolumn{1}{c}{$k_{F}$}  &
\multicolumn{1}{l}{ ${\cal E}/A$} &
 $\tilde{M}/M$ &
\multicolumn{1}{c}{ $U_{S}$} &
\multicolumn{1}{c}{ $U_{V}$} &
\multicolumn{1}{c}{ $\kappa$} & &
 ${\cal E}/A$  \\
\multicolumn{1}{c}{(fm$^{-1}$)} &
 (MeV) &   & (MeV) & (MeV) & (\%)& & (MeV) \\ \\
 \hline \\
0.8 &  --6.80 & 0.851 & --140.3 & 108.6 & 23.0 && --7.21\\
0.9 &  --8.30 & 0.812 & --176.6 & 137.2 & 19.1 && --8.79\\
1.0 &  --9.69 & 0.773 & --212.9 & 165.8 & 16.9 && --10.20\\
1.1 & --10.64 & 0.732 & --252.1 & 197.5 & 13.6 && --11.15 \\
1.2 & --11.57 & 0.688 & --292.8 & 229.8 & 13.0 && --12.46 \\
1.3 & --12.25 & 0.644 & --334.0 & 262.8 & 13.8 && --13.87 \\
1.35 & --12.24 & 0.620 & --356.6 & 281.8 & 14.5 && --14.35 \\
1.4 & --11.99 & 0.601 & --374.2 & 296.4 & 15.3 && --15.02 \\
1.5 & --10.06 & 0.561 & --411.8 & 330.9 & 16.0 && --15.73 \\
1.6 &  --5.72 & 0.519 & --451.7 & 373.2 & 17.5 && --15.65 \\
1.7 &  +1.81 & 0.482 & --486.3 & 416.6 & 20.2 && --15.01 \\
1.8 &  13.50 & 0.450 & --516.8 & 464.6 & 23.7 && --13.50 \\
1.9 &  29.85 & 0.426 & --538.8 & 513.8 & 26.9 && --11.22 \\
2.0 &  52.48 & 0.409 & --554.6 & 567.7 & 29.2 &&  --7.39 \\
2.1 &  82.54 & 0.391 & --571.8 & 639.4 & 31.8 &&  --1.75 \\
2.2 & 119.89 & 0.379 & --583.0 & 717.7 & 34.9 &&  +5.88 \\
\\ \hline  \\
\end{tabular}
\end{table}

\begin{figure}[t]
\centerline{\psfig{figure=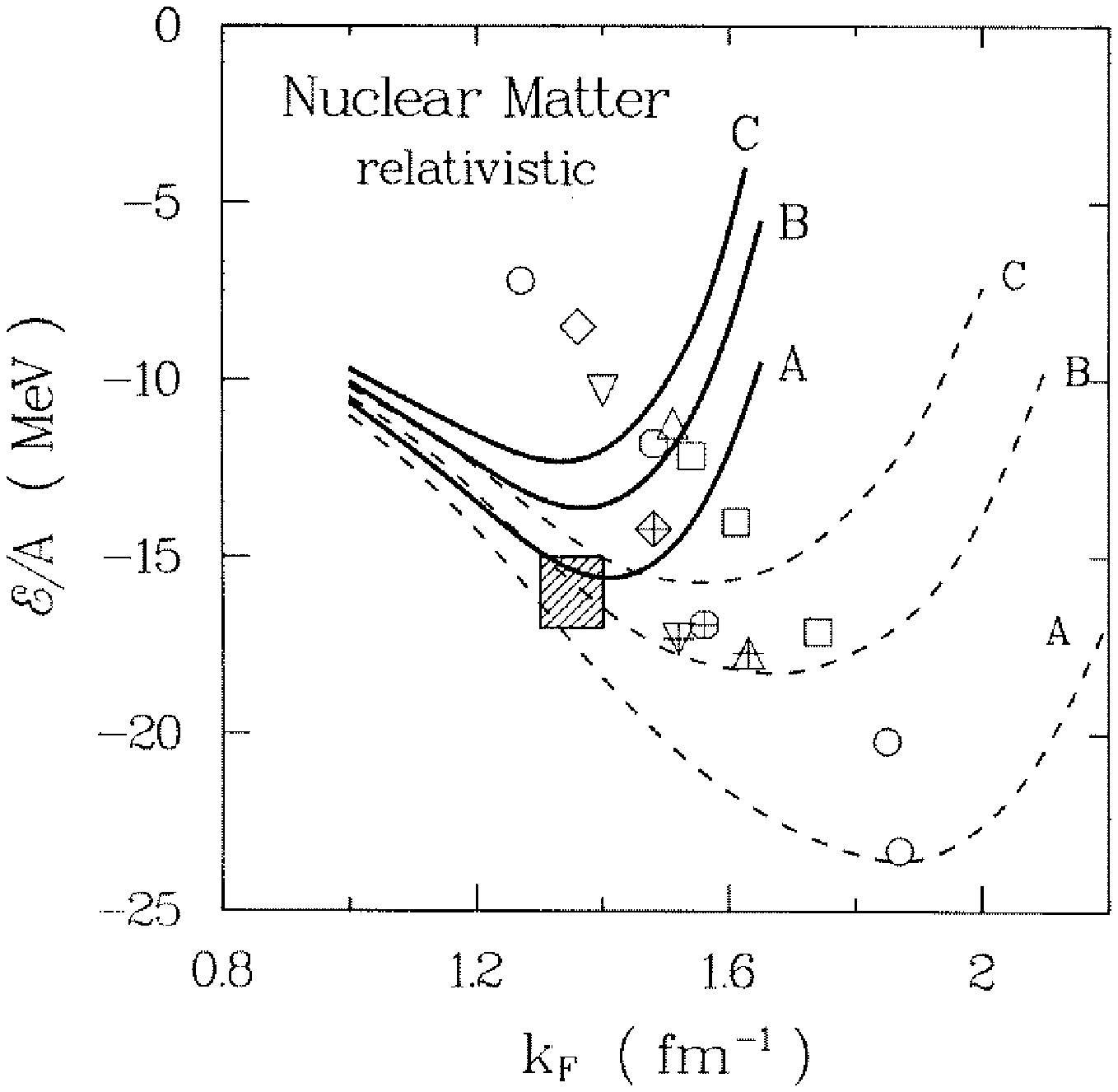,width=9cm}}
\caption{Results from calculations with a family of relativistic
potentials revealing a new Coester band which meets the
empirical area; solid lines: relativistic, dashed lines:
non-relativistic calculations. For comparison, saturation points
from conventional calculations are displayed in the background
(cf.\ Fig.~1). The shaded square denotes the empirical value
for nuclear matter saturation.}
\end{figure}

We apply now the three OBE potentials presented in Sect.~3.3 to nuclear
matter.
We stress again, that our potentials
use 
for $\pi NN$ the coupling of
derivative (pseudovector) type.
This is improtant because the pseudoscalar $\pi NN$ coupling
leads to an unrealistcally large attractive medium effect~\cite{MB85}.

The main difference between the three potentials
applied here, is the strength of the tensor force as
reflected in the predicted D-state probability of the deuteron, $P_D$.
With $P_D=4.5\%$, Potential A
has the weakest tensor force. Potential B and C predict 5.1\% and 5.5\%,
respectively (cf.\ Table~2).
 It is well-known \cite{Mac89} that the strength of the tensor
force determines the location of the nuclear matter saturation point on the
Coester band \cite{Coe}. To find out the structure of the Coester band,
predictions from more than one potential are needed.

All results presented in this section are obtained either in
the Bruckner-Hartree-Fock (BHF) or the Dirac-Brueckner-Hartree-Fock (DBHF)
approximation; as mentioned before,
occasionally, we will denote these two methods also as the
`non-relativistic' and the `relativistic' calculation, respectively.

The repulsive relativistic effect in nuclear matter as created by the DBHF
method in shown in Fig.~3. In addition, in Table 3, we list the following
quantities as a function of the Fermi momentum $k_{F}$: the energy per
nucleon, ${\cal E}/A$, the ratio $\tilde{M}/M$,
 the single-particle scalar and vector
potentials $U_{S}$ and $U_{V}$, and the wound integral $\kappa$. (For
the definition of $\kappa$ and for
explicit formulae appropriate for the momentum-space framework,
 see section 5 of Ref.~\cite{HT70}.)
$\kappa$ can be understood as the expansion
parameter for the hole-line series.

As mentioned in Sect.~4,
the suppression of the $\sigma$ contribution can be understood
in simple terms by considering
the covariant one-$\sigma$-exchange amplitude,
  for ${\bf q'}={\bf q}$ and $\lambda_{i}=\lambda_{i}'$
(as used in the Hartree approximation), in which case,
due to the covariant normalization
 of the Dirac spinors, the numerator
becomes ${\bf 1}$. Since the physical states of the nucleons in nuclear 
matter, $w$,
are to be normalized by $w^{\dagger} w = 1$ implying
$w\equiv\sqrt{\tilde{M}/\tilde{E}}\times\tilde{u}$,
 the sigma (as any other) contribution
gets the (scalar density) factor $(\tilde{M}/\tilde{E})^{2}$ (see second term on
the
r.h.s.\ of Eq.\ (99)) which decreases with decreasing $\tilde{M}$
(i.\ e.\ increasing density).
A corresponding consideration for the time-like ($\gamma_{0}$) component
of $\omega$-exchange would lead to no changes for that contribution.
However, due to the exchange term and correlations there is a small
enhancement of the
repulsion created by the $\omega$ with density.

>From the numbers given in Table 3 it is seen that the relativistic
effect on the energy per nucleon, $\Delta ({\cal E}/A)_{rel}$
(i.\ e.\ the difference between the relativistic and non-relativistic
calculation),
is well fitted by the ansatz
\begin{equation}
\Delta({\cal E}/A)_{rel} \approx  2 \mbox{ MeV} \times (\rho/\rho_{0})^{8/3},
\end{equation}
which is suggested by an estimate by Keister and Wiringa \cite{KW}.

Furthermore, Table 3 shows that up to nuclear matter density the wound
integral $\kappa$ is slightly smaller for the relativistic approach than for
the non-relativistic one. This implies that in this region the relativistic
many-body scheme should be slightly better convergent. Beyond nuclear matter
density, the situation is reversed. In addition, it is amusing to note that
for all values of $k_{F}$ the ratio $\tilde{M}/M$ is almost the same for
the non-relativistic and the relativistic
approach. Low values for $\tilde{M}/M$ have often been critizised. However,
they are not a consequence of the relativistic approach but are due to the
Brueckner-Hartree-Fock approximation. Higher order corrections will enhance
this quantity. For a recent discussion of the effective mass in relativistic
and non-relativistic approaches see Refs.~\cite{JM,LMB92}.

The representation of nucleons by Dirac spinors
with an effective mass, $\tilde{M}$, can be interpreted, as taking
virtual nucleon-antinucleon excitations
in the many-body environment
(many-body Z-graphs)
 effectively into account \cite{Bro87}.
This can be made plausible by expanding the spinors Eq.\ (92)
in terms of (a complete set of) spinor solutions of
the free Dirac equation which
will necessarily also include solutions representing negative
energy (antiparticle) states \cite{Ana}.

In Table 4, we compare
the contributions in various partial-wave states as obtained in a relativistic
calculation to that from the corresponding non-relativistic one.
Detailed investigations have shown that the  repulsive
relativistic effect
seen in that table
for the $P$-wave contributions is essentially due to $\sigma$
suppression
together with a  signature of  spin-orbit force enhancement.
The change of the  $^{1}\!S_{0}$ contribution
 is so small, because of a cancelation
of effects due to $\sigma$ and $\rho$. Apart from $\sigma$ reduction,
the repulsive
effect in $^{3}\!S_{1}$ is due to a suppression of the
twice iterated one-pion exchange for reasons quite analogous to
the sigma suppression.

A comparison between relativistic and non-relativistic Brueckner-Hartree-Fock
calculations for all three potentials is shown in Fig.~4. For the non-
relativistic approach, the three saturation points are clearly on the Coester
band. However, using the relativistic method, the saturation points are
located on a new band which is shifted towards lower Fermi momenta (densities)
and even meets the empirical area. This is a very desirable effect. The reason
for this shift of the Coester band is the additional strongly density
dependent repulsion which
 the relativistic approach gives rise to. In Table 5, the
saturation points for the different potentials are given for the relativistic
as well as the non-relativistic calculation. In the relativistic case, the
incompressibility of nuclear matter using Potential B is about 250 MeV which
is in satisfactory agreement with the empirical value of $210\pm 30$ MeV
 \cite{Bla}.
Note that in the relativistic Walecka model, 540 MeV is obtained for the
compression modulus \cite{SW}.

For completeness, we present in Table 6 the Landau parameters at various
densities as obtained in a relativistic as well as a non-relativistic
nuclear matter calculation.
Based on the nuclear matter G-matrix, the effective particle-hole interaction
at the Fermi surface is calculated and, multiplied by the density of
states
$k_{F}M/(\hbar^{2}\pi^{2})$,  parametrized by:
$F=f+f'\tau_{1}\cdot\tau_{2}+g\sigma_{1}\cdot\sigma_{2}
+g'\sigma_{1}\cdot\sigma_{2}\tau_{1}\cdot\tau_{2}$.
>From an expansion of the parameters in terms of Legendre polynominals,
$P_{l}$,
we give in Table 6 the coefficient for $l=0$. For more details and the
empirical values see Ref.~\cite{KNS}.
Notice that the prediction for the Landau parameter $f_{0}$ is considerably
improved in the relativistic calculation
 without deteriorating the other parameters.
Sum rules require $f_{0} > -1$ at nuclear matter
density \cite{Mig67}.
Finally, Tables 7 and 8 contain more results for
Potentials A and C.

Concerning nuclear matter 
at higher densities
and neutron matter, 
we like to refer the interested reader to
Ref.~\cite{LMB92}.

\section{In-Medium NN Cross Sections}

\begin{figure}[t]
\centerline{\psfig{figure=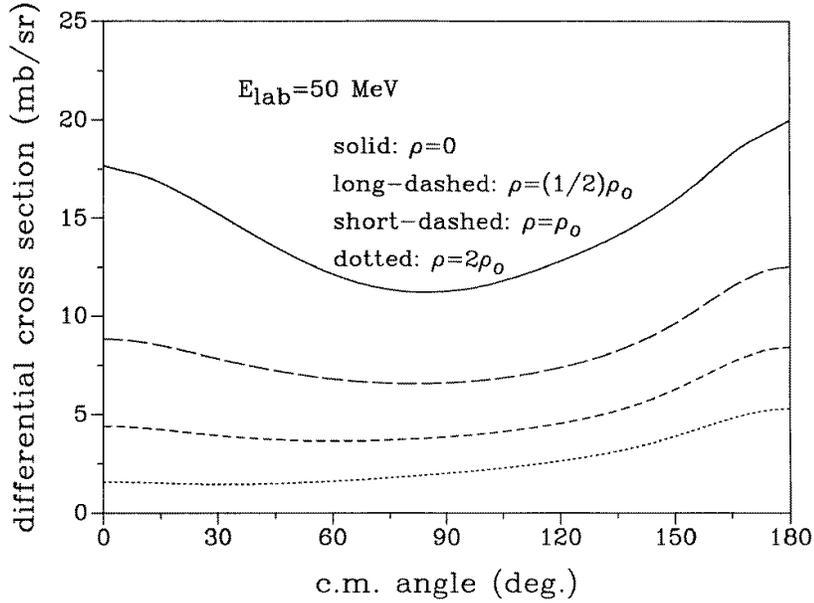,width=11cm}}
\caption{In-medium $np$ differential cross sections at 50 MeV
laboratory kinetic energy, as obtained for various densities.}
\end{figure}

\begin{figure}[t]
\centerline{\psfig{figure=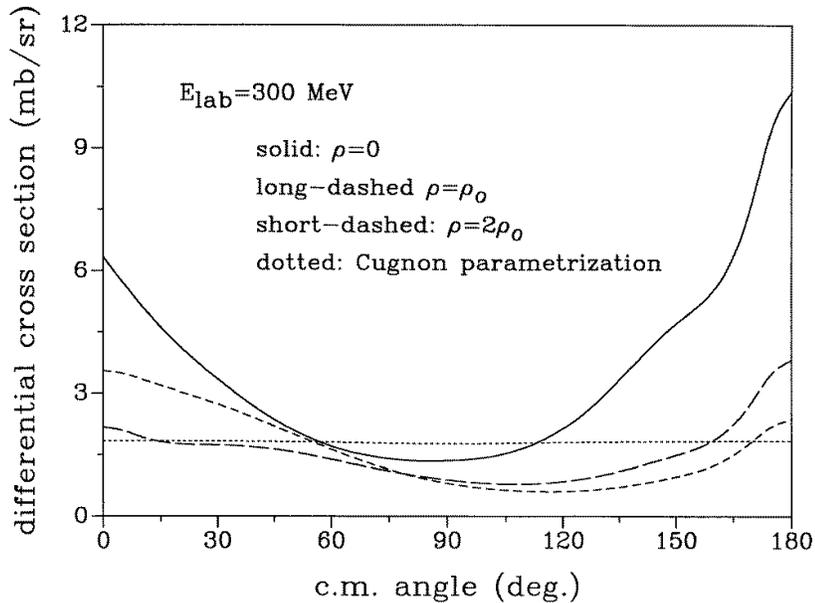,width=11cm}}
\caption{In-medium $np$ differential cross sections at 300 MeV
laboratory kinetic energy, as obtained for various densities
using the Bonn~A potential (solid, long-dashed, and short-dashed lines). 
The dotted line shows the parametrisation by Cugnon {\it et al.} [92].}
\end{figure}

\begin{figure}[t]
\centerline{\psfig{figure=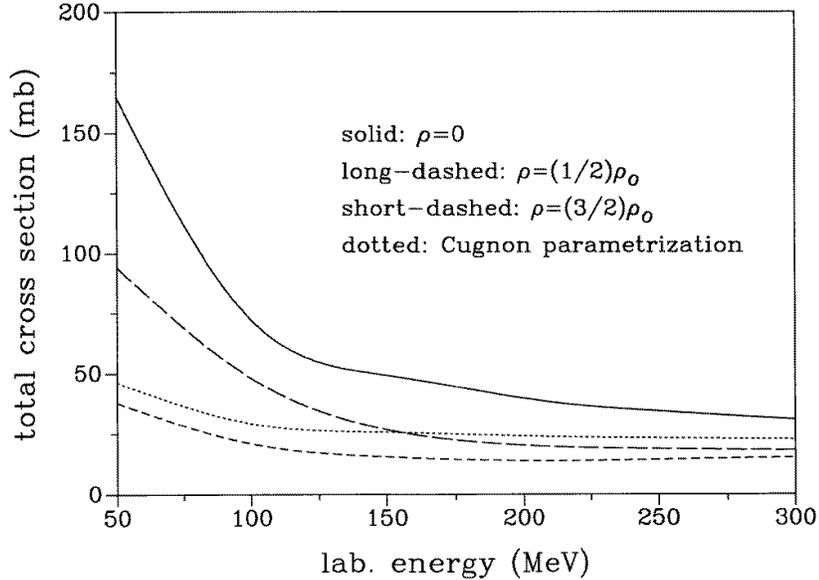,width=11cm}}
\caption{In-medium $np$ total cross sections
as obtained for various densities
using the Bonn~A potential (solid, long-dashed, and short-dashed lines). 
The dotted line shows the parametrisation by Cugnon {\it et al.} [92].}
\end{figure}

The investigation of 
in-medium NN scattering is of  interest
for intermediate-energy heavy-ion reactions. 
Experimentally, nucleus-nucleus collisions at intermediate energies
provide a unique opportunity to form  a piece of nuclear 
matter in the laboratory 
with a density up to 2-3$\rho _0$ (with $\rho _0$, in the range of 0.15 to
0.19 fm$^{-3}$,
the saturation density of normal nuclear matter; in this section we
use $\rho _0$=0.18 fm$^{-3}$) \cite{stock,grei}.
Thus it is  possible to study the properties of hadrons in
a dense medium. Since this piece of dense nuclear matter exists only
for  a very short  time (typically $10^{-23}$
-$10^{-22}$ s), it is necessary
to use transport models to simulate the entire collision  process and
to deduce the properties of the intermediate stage from the known initial
conditions and  the final-state observables. At intermediate energies,
both
the mean field and the two-body collisions play an equally important role
in the dynamical evolution of the colliding system; they have to be 
taken into account in the transport models on an equal footing, 
together with a proper treatment of the Pauli blocking for the 
in-medium two-body collisions. The Boltzmann-Uehling-Uhlenbeck (BUU)
equation \cite{bert1,mosel} 
and quantum molecular dynamics (QMD)
\cite{aich3,aich4}, as well as their relativistic extensions 
(RBUU and RQMD)
\cite{ko2,cass1,sorge,maru}, are promising transport models for
the description of intermediate-energy heavy-ion reactions. In 
addition to the mean field~\cite{LM93a}, 
the in-medium NN cross sections are also 
important ingredients of these transport models. Specifically, 
in-medium total as well as differential NN cross sections are needed
by these models in dealing with the in-medium NN scattering
using a Monte Carlo method. 

In this section, we will
calculate the elastic in-medium NN 
cross sections in a microscopic way.
We base our investigation on
the Bonn potentials presented in Sect.~3
and the Dirac-Brueckner approach 
explained in Sect.~4.
The original, detailed paper on this subject is Ref.~\cite{LM93}.

In-medium NN cross sections can be  calculated from the 
in-medium $\tilde 
G$ matrix. 
Here, we evaluate the $\tilde G$-matrix
in the
center-of-mass (c.m.) frame of the two interacting nucleons, 
i.~e., we use Eq.~(95) with ${\bf P} = 0$.
For the starting energy, Eq.~(96), we have now: $\tilde z =
2 \tilde E_q = 2 \sqrt{\tilde m^2 + q^2}$, where $q$ is related to
the kinetic energy
of the incident nucleon in the  ``laboratory system" ($E_{lab}$),
in which the other nucleon is at rest, 
by: $E_{lab}=2 q^2/m$.
Thus, we consider two colliding nucleons in nuclear matter.
The Pauli projector is represented by
one Fermi sphere as in conventional nuclear matter calculations.
This
Pauli-projector, which is originally defined in the nuclear matter rest
frame, must be boosted to the c.m. frame of the two interacting nucleons. For
a detailed discussion of this and the explicit formulae, see
Refs. \cite{HM87,HS87}.

In Fig. 5 we show the results for the in-medium $np$ differential cross sections
as a function of the c.m. angle 
at  $E_{lab}$=50 MeV and in Fig.~6 at 300 MeV. 
We consider the medium densities
$\rho$=0 (free-space scattering),
(1/2)$\rho_0$,
$\rho_0$, and
$2\rho_0$.  The results are obtained by using the Bonn
A potential. At low incident energies (Fig.~5),
the $np$ differential cross section always
decreases with increasing density, at both forward and backward
angles.
At high incident energies (Fig.~6), the {\it np} differential
cross section for  forward angles  decreases when going
from $\rho$=0 to $\rho _0$ and then increases for higher 
densities. The differential
cross section at  backward angles always decreases with density.
While the free {\it np}
differential cross sections are  highly anisotropic, 
the in-medium cross sections become more isotropic with increasing 
density.

We have also calculated the in-medium cross sections using other
potentials (Bonn~B and C) and found very little difference as compared
to Bonn~A. Thus, there is fortunately little model dependence
in our predictions.

It is also interesting to compare the 
results obtained in this work with the  parametrized NN cross sections
proposed by
Cugnon {\it et al.} \cite{bert1,cugn1} which are  often used in 
transport models such as BUU and QMD. This comparison is 
included in Fig.~6.
It is clearly seen that, while the Cugnon parametrization
is almost isotropic,
the microscopic results still have some  anisotropy at all densities
considered. The anisotropy
in the present results decreases with 
increasing density.
 There is also density dependence
in the microscopic differential cross section, 
while  the 
Cugnon parametrization is density independent.  
 
We mention that Cugnon {\it et al}. \cite{cugn1} have parametrized the free $pp$
cross sections. This explains the almost-isotropy in their 
differential cross sections as well as the lack of the density
dependence. 
Note that in the work of Cugnon {\it et al.,} \cite{cugn1}, no difference
is made between proton and neutron; thus, the $pp$ cross sections are also
used for $np$ scattering. There are, however, well-known differences
between $pp$ and $np$ cross sections which in the more accurate 
microscopic calculations of the near future may be relevant. The
difference between the (in-medium) $np$ and $pp$ cross sections is
discussed in detail in Ref.~\cite{LM94}. In this section, we restrict
our discussion of
``NN cross sections" to $np$ cross sections.
 
The {\it np} differential cross section in  free space can be well
parametrized by the following simple expression:
\begin{equation}
{d\sigma \over d\Omega}(E_{lab}, \theta )={17.42\over 1.0+0.05(E_{lab}
^{0.7}-15.5)}exp[b(cos^2\theta+sin^2{\theta \over 7}-1.0)]
\end{equation}
with
$$
b=0.0008(E_{lab}^{0.54}-4.625) ~{\rm for} ~~E_{lab}\leq 100 {\rm MeV}
$$
and 
$$
b=0.0006(36.65-E_{lab}^{0.58}) ~{\rm for} ~~E_{lab}> 100 {\rm MeV}.
$$
with $E_{lab}$ in the units of MeV.

It would be useful to parametrize the in-medium 
$np$ differential cross section as well. However, the complicated
dependence of the in-medium differential cross sections on angles,
energy, and especially density makes this very difficult.
Instead, we have prepared a data file, containing 
in-medium differential cross section 
as a function of angle for a number of densities 
and energies, from which the differential cross sections  
for all densities in the range 0-3$\rho _0$ and all energies in the range 0-300
MeV can be interpolated.
This data file is available from one of the authors (R.\ M.) upon request.

In addition to the in-medium NN differential cross  sections which enter the 
transport models to determine the direction of the outgoing nucleons,
the in-medium total NN cross sections, $\sigma _{NN}$, are also 
of interest.  They provide a  criterion for  whether a pair of
nucleons will collide or not by comparing their closest distance to 
$\sqrt {\sigma _{NN}/\pi }$. 
We show in Fig.~7 the in-medium total
cross sections as function of the incident energy $E_{lab}$ and
density. 
It is seen that the in-medium total
cross sections decrease substantially with  increasing 
density, particularly for low incident energies.

Figure~7 also includes a comparison with the total cross sections
used by
Cugnon {\it et al.}~\cite{bert1,cugn1} (dotted line in Fig.~7). 
It is seen that  at low  energies and
low  densities,  the Cugnon parametrization underestimates the
microscopic results, while at higher energies and higher density
it is the other way around.  Note that the Cugnon parametrization
is not density dependent, and, thus, predicts the same for all densities.

Finally,  we propose a parametrization for the total $np$ cross section as
a function of the incident energy $E_{lab}$ and density $\rho$
\begin{equation}
\sigma _{np}(E_{lab},\rho )=(31.5+0.092{\rm abs}(20.2-E_{lab}^{0.53})^{2.9})
{1.0+0.0034E_{lab}^{1.51}\rho ^2\over 1.0+21.55\rho ^{1.34}}
\end{equation}
where $E_{lab}$ and $\rho$ are in the units of MeV and fm$^{-3}$, 
respectively.

The major conclusions of the present microscopic
calculations are: 
\begin{itemize}
\item
There is strong density dependence for the in-medium cross
sections. With the increase of density,
the cross sections  decrease. This indicates that 
a proper treatment of the density-dependence of the in-medium NN cross sections
is important.  
\item
Our microscopic predictions  differ from the  commonly used  
parametrizations of the differential  and the total 
cross sections developed by Cugnon {\it et al}.
\cite{bert1,cugn1}. The Cugnon parametrization
underestimates the anisotropy of the in-medium
{\it np} differential cross sections.
In the case of  the total cross
sections, the Cugnon parametrization 
either underestimates or overestimates the microscopic 
results, depending on energy and  density. 
\end{itemize}
More details of this investigation can be found in Refs.~\cite{LM93,LM94}.

\section{Finite Nuclei}

Encouraged by the good results for nuclear matter, one may now try to 
describe finite nuclei
starting from the free-space
nucleon-nucleon interaction. A straightforward way
would be to solve the relativistic Brueckner-Hartree-Fock equations for
finite systems. This is, however, an extremely difficult 
task~\cite{MMB90,FMM93}. 
Therefore,
it might be 
a reasonable next step to incorporate the DBHF results in
a relativistic Hartree framework where the coupling constants are made
density dependent so as to reproduce the nuclear matter results. This
relativistic density dependent Hartree (RDDH) approach is similar in spirit
to the work by Negele in the nonrelativistic approach~\cite{Neg70}.
     
The working basis of the RDDH approach for finite nuclei is the
relativistic Hartree Lagrangian~\cite{Bro} (sigma-omega model Lagrangian
of Walecka~\cite{Wal74}). Writing explicitly the density dependence of the
coupling constants,we have
\begin{eqnarray}\label{1}
\cal{L}_{\hbox{\scriptsize RDDH}} & = & \bar \psi\left(i\gamma_{\mu}\partial^{\mu}
- M - g_{\sigma}(\rho)\sigma - g_{\omega}(\rho)\gamma_{\mu}\omega^{\mu}
\right)\psi \nonumber\\
& & + \frac{1}{2}\left(\partial^{\mu}\sigma\right)^{2} -
\frac{1}{2}m_{\sigma}^{2}\sigma^{2} -
\frac{1}{4}\left(\partial_{\mu}\omega_{\nu}-\partial_{\nu}\omega_{\mu}
\right)^{2} + \frac{1}{2}m_{\omega}^{2}\omega_{\mu}^{2} \;,
\end{eqnarray}
in conventional notation~\cite{BD64}.
In Hartree approximation (mean field approximation), the
nucleon self-energy in nuclear matter is given by
\begin{eqnarray}\label{2}
\Sigma_{\hbox{\scriptsize RDDH}}(\rho) & = & \hbox{U}_{s}(\rho) +
\hbox{U}_{v}(\rho)\gamma_{0} \;.
\end{eqnarray}
Here, the scalar and the vector potentials are expressed in terms of the
coupling constants g$_\sigma(\rho)$ and g$_\omega(\rho)$ through
\begin{eqnarray}\label{3}
\hbox{U}_{s}(\rho) & = & -  \frac{{g_{\sigma}}^2(\rho)}{m_{\sigma}^{2}}
\rho_{s} \nonumber\\
\hbox{U}_{v}(\rho) & = & \frac{{g_{\omega}}^2(\rho)}{m_{\omega}^{2}}
\rho_{v} \;,
\end{eqnarray}
where $\rho_s$ and $\rho_v$ are the scalar and the vector densities,
respectively:
\begin{eqnarray}\label{4}
\rho_{s} & = & <\bar \psi \psi> \;=\; 4\int\limits_{0}^{k_{\hbox
{\scriptsize F}}}
\frac{d^{3}k}{(2\pi)^{3}}\,\frac{M^{*}}{E^*}\; , \nonumber\\
\rho_{v} & = & <\bar \psi \gamma_{0} \psi>
\;=\; 4\int\limits_{0}^{k_{\hbox{\scriptsize F}}}
\frac{d^{3}k}{(2\pi)^{3}} \; .
\end{eqnarray}
The scalar potential is related to the effective mass by
$M^{*}\, =\, M + \hbox{U}_{s}(\rho)$.
The connection of the RDDH approach
to the DBHF theory is made through the nucleon self-energy in nuclear
matter, Eq.~(109). 
In fact, one can express the DBHF self-energy in nuclear matter
in this form \cite{BM} using the symmetry requirements and
redefinition of various terms through the use of the Dirac equation
\cite{Jam83}. The density dependent coupling constants are then obtained
through Eqs.~(110) and (111), where U$_s(\rho)$ and U$_v(\rho)$ are the
results of the DBHF calculations of nuclear matter.
     
We write the equations of motion for finite nuclei for completeness. The
normal modes of the nucleon are calculated with the Dirac equation,
\begin{eqnarray}\label{5}
\left(-i
\mbox{\boldmath $\alpha \cdot \nabla$} + \beta M^{*}(r) + V(r) \right)
\psi_{i}({\bf r}) & = & E_{i}\psi_{i}({\bf r})
\end{eqnarray}
with $M^{*}(r)\,=\,M+g_{\sigma}(r)\sigma(r)$ and $\hbox{V}(r)\,=\,
g_{\omega}(r)\omega^{0}(r) + e\frac{1-\tau_{3}}{2}A^{0}(r)$.

The Klein-Gordon equations for $\sigma$, $\omega^0$, and $A^0$ are
\begin{eqnarray}\label{6}
\left(-\triangle + m_{\sigma}^{2}\right)\sigma(r) & = & - g_{\sigma}(r)
\rho_{s}(r) \nonumber\\
\left(-\triangle + m_{\omega}^{2}\right)\omega^{0}(r) & = &   g_{\omega}(r)
\rho_{v}(r) \nonumber\\
-\triangle A^{0}(r) & = & e \rho_{p}(r) ,
\end{eqnarray}
where $\rho_s$, $\rho_v$ and $\rho_p$ are the scalar, vector and proton
densities, respectively, which are obtained through the Dirac wavefunctions.
We solve the coupled differential equations self-consistently. The center
of mass corrections are applied in the same way as in Ref.~\cite{Rei86}.

\begin{figure}[t]
\centerline{\psfig{figure=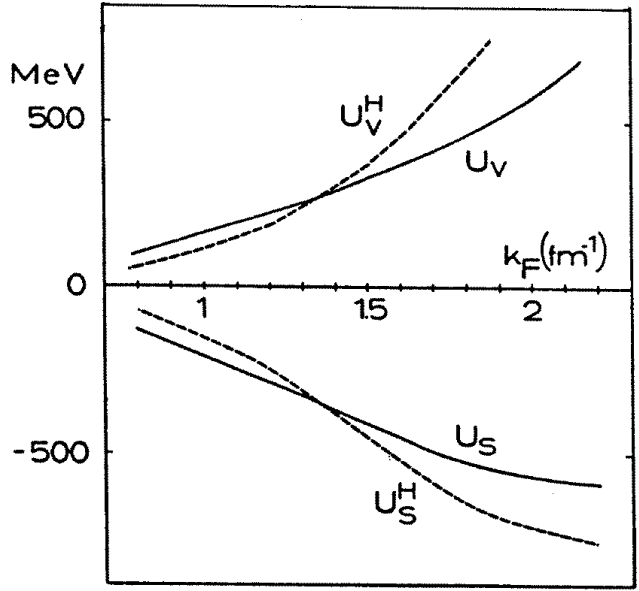,width=8.5cm}}
\caption{Vector and scalar potentials for relativistic density dependent
Hartree (full lines) and relativistic Hartree only (dashed lines) as a
function of k${_F}$.}
\end{figure}

In Fig.~8, the
vector and the scalar potentials, U$_v$ and U$_s$ are depicted as a
function of k$_F$. At normal matter density, k$_F$ = 1.35 fm$^{-1}$,
U$_v$ = 274.7 MeV and U$_s$ = --355.7 MeV. As a comparison, the vector and
scalar mean-field potentials are shown using constant (density independent)
couplings; i.e. the scalar and vector potential parameters are the ones of
Potential A.
This comparison clearly indicates that one needs density dependent
coupling constants 
in order to reproduce the nuclear matter results
within the relativistic Hartree framework.
     
Now we fit the coupling constants g$_\sigma(\rho)$ and g$_\omega(\rho)$ to
the self-energies U$_s$ and U$_v$ with the choice that m$_\sigma$ and
m$_\omega$ are the masses in Potential A as a standard choice;
m$_\sigma$ = 550 MeV and m$_\omega$ = 782.6 MeV. As can be seen 
from Eq.~(110), the relevant quantities are g$_\sigma^2$ and 
g$_\omega^2$. At k$_F$ = 0.8, 1.1, 1.5 
we obtain g$_\sigma^2$/4$\pi$ = 12.3,
8.91, 6.23 and g$_\omega^2$/4$\pi$ = 18.63, 13.48, 9.06, respectively. This 
shows that the coupling constants at the surface are more than 40\% bigger 
than in the interior.
At 
nuclear matter density (k$_F$ = 1.35fm$^-1$) the coupling constants are not 
modified as one can see from Fig.~8. But at smaller densities 
both coupling constants are growing. 
We note
here that U$_s$ and U$_v$ are, in principle, dependent not only on the
density but also on the momentum of the nucleon. 
We neglect this weak momentum dependence in our study of finite
nuclei.
     
We calculate $^{16}$O and $^{40}$Ca as examples for finite nuclei within the
RDDH approach. The calculated results on the binding energies, single
particle energies and charge radii (using the Bonn A potential) 
are tabulated in Table 9. 
For comparison,
results of nonrelativistic Brueckner-Hartree-Fock calculations
for $^{16}$O with the same potential (Bonn A) are shown in the column N-BHF
\cite{FMM93}. It is interesting
to find that the RDDH results are very close to
experiment. The improvements, as compared to N-BHF, are remarkable. The root
mean square charge radius
is almost perfect, while the binding energy is slightly smaller than
experiment. If we compare relativistic and nonrelativistic results, we find 
that the radius of the relativistic calculation is larger. This is natural 
since, e.~g., 
the lower component of the relativistic 1p3/2 - wavefunction looks 
like a nonrelativistic 1d3/2 - wavefuction. This shifts part of the density 
to the surface and leads to a larger radius in the relativistic case. This 
has also consequences for the binding energy per nucleon. Using the 
Bethe-Weizsaecker mass formula, we can estimate that surface and Coulomb 
effects lead to 2 to 3 MeV less repulsion for the relativistic calculation 
since the radius is larger. Although the volume effect is 1 to 2 MeV more 
repulsive in the relativistic case which has been determined in nuclear 
matter in Ref.~\cite{BM90}, the relativistic description of 
finite nuclei yields altogether more binding energy per nucleon. This shows 
that relativistic effects lead off the Coester band, which exists for 
finite nuclei, too.

\begin{table}[t]
\caption{The binding energy per nucleon,
root-mean-square charge radius and single particle energies are displayed
for the relativistic density dependent Hartree(RDDH) approach, the
nonrelativistic Brueckner-Hartree-Fock(N--BHF) method and experiment,
respectively. The Bonn A potential is used.
The upper part of the table is for $^{16}$O and the lower one
for $^{40}$Ca.}
\begin{center}
\begin{tabular}{|c||c|c||c|}\hline
$^{16}O$  & RDDH & N--BHF & Experiment \\ \hline \hline
BE/A [MeV] & --7.5 &  --5.95  & --7.98  \\ \hline
r$_c$ [fm]& 2.66 & 2.31  &$ 2.70\pm0.05 $ \\ \hline
$\epsilon$(1s$_{1/2}$) [MeV]& --43.5 & --56.6  &$ -40\pm8 $ \\ \hline
$\epsilon$(1p$_{3/2}$) [MeV]& --21.8 & --25.7  & --18.4  \\ \hline
$\epsilon$(1p$_{1/2}$) [MeV]& --16.5 & --17.4  & --12.1  \\ \hline
\end{tabular}
\end{center}
\begin{center}
\begin{tabular}{|c||c|c||c|}\hline
$^{40}Ca$ & RDDH & N--BHF & Experiment \\ \hline \hline
BE/A [MeV] & --8.0 & --8.29 & --8.5  \\ \hline
r$_c$ [fm]& 3.36 & 2.64 & 3.5  \\ \hline
$\epsilon$(1s$_{1/2}$) [MeV]& --53.3 & -- &$ -50\pm8 $ \\ \hline
$\epsilon$(1p$_{3/2}$) [MeV]& --36.0 & -- &$ -34\pm5 $ \\ \hline
$\epsilon$(1p$_{1/2}$) [MeV]& --32.5 & -- &$ -34\pm5 $ \\ \hline
$\epsilon$(1d$_{5/2}$) [MeV]& --19.3 & --30.2 &$ -14\pm2 $ \\ \hline
$\epsilon$(2s$_{1/2}$) [MeV]& --14.3 & --24.5 &$ -10\pm1 $ \\ \hline
$\epsilon$(1d$_{3/2}$) [MeV]& --13.6 & --16.5 &$  -7\pm1 $ \\ \hline
\end{tabular}
\end{center}
\end{table}

\begin{figure}[t] 
\centerline{\psfig{figure=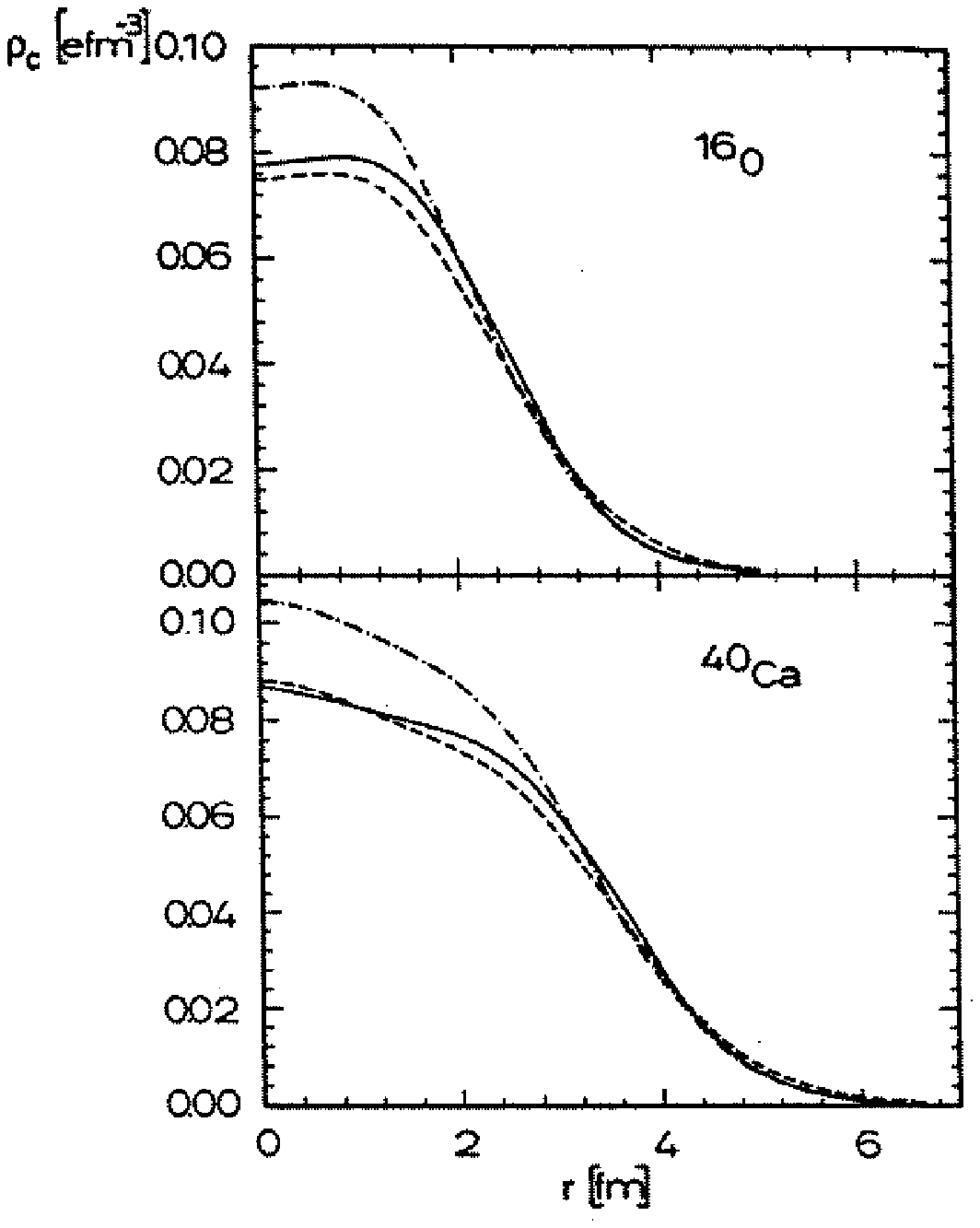,width=6.5cm}}
\caption{Comparison of the densities in relativistic density dependent Hartree
and experiment for $^{16}$O and $^{40}$Ca, respectively. The results 
with the Bonn A potential are shown by the dash-dotted 
curve, while the corresponding ones for Bonn C are depicted by the dashed 
one.}
\end{figure}

In Fig.~9, we show a comparison with experimental 
charge density distributions obtained
from elastic electron scattering~\cite{Rei86}. The RDDH
results with the Bonn A potential 
(dash-dotted curve) compare fairly well with the 
experimental data.
Looking more closely, the
central density comes out to be higher than experiment and the
density falls off slightly faster than experiment. This observation of the
density and the slightly smaller binding discussed above, seems to be the
reflection of the equation of state of nuclear matter as shown in Fig.~4.
In fact, when we take the DBHF results from the Bonn C potential,
whose saturation density
is almost perfect but the saturation energy is about 4 MeV above the
experimental value, the charge density distributions are found to be very
good as can be seen in Fig.~9 (dashed curve). In this case, however, the 
binding energies
of finite nuclei are found to be too small; i.e. E/A = - 5.9 MeV for $^{16}$O
and E/A = - 6.0   MeV for $^{40}$Ca.

Another issue is the density-dependence of meson masses; this is
discussed in Ref.~\cite{BT90}.

\section{Summary and Outlook}

In this chapter we have presented the formalism for a
relativistic approach to two-nucleon scattering and nuclear structure.
The latter is based upon a relativistic extension of Brueckner theory.
The essential idea of this approach is to use  the Dirac equation for the
single particle motion. The nucleon self-energy in nuclear matter obtained in
this framework consists of a large (attractive) scalar and (repulsive) vector
field. The size of these potentials (several hundred MeV) motivates the use of
the Dirac equation.

Furthermore,
we have applied
relativistic meson-exchange potentials appropriate for this
approach.
The one-boson-exchange (OBE) model includes the six non-strange bosons
with masses below 1 GeV/c$^2$, $\pi,\eta,\sigma,\delta,\omega,$ and $\rho$.
The potentials describe low energy NN scattering and the properties of the
deuteron quantitatively. Thus, they are  suitable for (parameter-free)
microscopic nuclear structure calculations.
 Apart from the usual interaction Lagrangians for heavy mesons,
these potentials apply the pseudovector (gradient) coupling for the $\pi NN$
(and $\eta NN$) vertex. For a relativistic approach, it is necessary to apply
this coupling for the pseudoscalar mesons, since the commonly used
pseudoscalar coupling leads to unrealistically attractive contributions.
This has its origin in chiral symmetry.

The consideration of the two-nucleon system is based upon the relativistic
 three-dimensional reduction of the Bethe-Salpeter equation introduced by
Thompson. We derive and present the Thompson equation for an arbitrary frame. 
In two-nucleon scattering, this equation is applied in the two-nucleon c.m. 
system, while in nuclear matter, the nuclear matter rest frame is used. Thus, 
we calculate the nuclear matter $G$-matrix directly in the nuclear matter rest 
frame in which it is needed to obtain the energy of the many-body system. The 
advantage of this method is that, in the nuclear medium,
 the rather involved and tedious 
transformation between the two-nucleon c.m. frame and the nuclear matter rest 
frame is avoided.

Using  the
framework outlined, the properties of nuclear matter are calculated in
the   Dirac-Brueckner-Hartree-Fock (DBHF) approximation.
The single particle scalar and vector fields as well as the single particle
wave functions (Dirac spinors) are determined fully self-consistently. The
strong attractive scalar field leads to a reduction of the nucleon mass in the
medium, which  increases with density. One consequence of this is a
suppression of the (attractive) scalar-boson exchange. This strongly
 density-dependent effect improves nuclear saturation considerably such that
 the empirical saturation energy and density can be
explained correctly.
Our calculations include also the saturation mechanisms of conventional
 (non-relativistic) many-body theory (i.~e.\ Pauli
and  dispersion effects).
 The predictions for the compression modulus of nuclear
matter as well as the Landau parameters are in satisfactory agreement with
empirical information.

In spite of the success of the present calculations, several critical
questions can be raised. First, there will be contributions of higher order in
the conventional hole-line expansion. For non-relativistic Brueckner theory,
Day found an increase in the binding energy per nucleon of 5--7 MeV from the
three- and four-hole line contributions.
Note that in the work of Day the standard choice (`gap' choice) for 
the single-particle potential is used (this is true for all results 
displayed in Table~1 and Fig.~1).. However, in the calculations of our work
the continuous choice for the single-particle potential, Eq.~(98), is applied.
In the two-hole line approximation, this choice leads to about 4 MeV more 
binding energy per nucleon 
as compared to the gap choice. Investigations by the Li\`{e}ge 
group \cite{Grang} suggest that a lowest order calculation with
the continuous choice effectively includes the 
three-hole line contributions. In the light of this result, the contributions 
of higher order in the hole-line expansion, missing in our work,
 may be believed to be small.

In a recent study, a special class of ring-diagrams has been summed up to
infinite orders using our relativistic G-matrix. Only a
minor change of the present DBHF result is found \cite{JMK} which, however,
further improves the saturation density.

Secondly, one may question the OBE model. The weakest part of that model is
the scalar isoscalar $\sigma$ boson. However, as mentioned, Dirac-Brueckner
calculations have also been performed with more realistic meson models, in
which the fictitious $\sigma$ is replaced by an explicit description of the
$2\pi$-exchange contribution to the NN interaction. In those calculations,
 it is found that the
relativistic effect comes out almost exactly the same
as in the OBE model \cite{MB}. This is not too surprising since it has been
known for a long time that the $2\pi$-exchange part of the nuclear potential
is well approximated by the exchange of a single scalar isoscalar boson of
intermediate mass.

One may also criticize that the present approach is restricted to
 positive-energy nucleons. At a first glance, this may appear rather
inconsistent in a relativistic framework. However, calculations done
for 
 two-nucleon scattering have shown that the
contributions from virtual anti-nucleon intermediate states
 (pair-terms) are extremely small (when
the pseudovector coupling is used for the pion) \cite{FT80,ZT81}.
Thus, the medium effects on this small contributions will be even smaller.
This has been confirmed by the relativistic $G$-matrix calculations
including antiparticle intermediate states by
Amorim and Tjon~\cite{AT92}.

Since higher orders in the hole-line expansion are attractive,
while corrections from the Dirac sea (of anti-nucleons)
have (as far as calculations exist)
yielded additional repulsion \cite{HS87}, cancelations may occur between
 many-body contributions missing in the present approach.
 Thus, it is not unlikely that DBHF  may ultimately turn out to
be a reasonable approximation to this very complex (relativistic) many-body
 problem.

We have also calculated cross sections for NN scattering
in the nuclear medium. We find strong density dependence for
the in-medium cross sections. The results from our microscopic
derivation differ substantially from simple cross section
parametrizations that are commonly used in heavy ion reaction 
calculations.
Our in-medium cross sections together with our momentum-dependent
mean field~\cite{LM93a}, which are both derived
on an equal footing, can now be used as input for consistent
transport model calculations of heavy ion reactions.

Moreover, we developed a relativistic density-dependent
Hartree approach for finite nuclei, where the coupling
constants of the relativistic Hartree-Lagrangian are made
density-dependent and are obtained from the relativistic 
Brueckner-Hartree-Fock results of nuclear matter.
The results on binding energies and root-mean-square
radii of $^{16}$O and $^{40}$Ca agree very well with
experiment.

Further applications of our approach to the nucleon mean free path
and proton-nucleus scattering are presented
in Refs.~\cite{LMZ93,Li93}.

On a more fundamental level,
 one may raise the critical question how serious and reliable the
meson model is. Notice
 that the relativistic effects obtained in Dirac approaches
are intimately linked to the meson model for the nuclear force.
In the meson-exchange model for the NN interaction, the isoscalar vector meson
$\omega$ and the isoscalar $\sigma$ boson provide the largest contributions.
In our many-body framework, they are responsible for the large effective
scalar and vector fields obtained for the single particle potential (nucleon
self-energy) in the many-body system. Assuming that quantum-chromodynamics
(QCD) is the fundamental theory of strong interactions, one of the greatest
challenges
we have to face is the simple question: why do these meson models work so well
--- for two as well as many nucleons?
Chiral symmetry may provide the key to the answer.

\vspace{.5in}

This work was supported in part by the U. S. National Science Foundation
under Grant No.~PHY-9211607 and by the Deutsche Forschungsgemeinschaft
(SFB 201).

\appendix

\end{document}